\documentclass[useAMS,usenatbib]{mn2e}

\usepackage{graphicx}
\usepackage{psfig}
\usepackage{makeidx,amsmath,mathrsfs}
\usepackage{verbatim}

\onecolumn

\newcommand\aj{AJ}
\newcommand\apj{ApJ}
\newcommand\apjl{ApJ}
\newcommand\apjs{ApJS}
\newcommand\aap{A$\&$A}
\newcommand\mnras{MNRAS}
\newcommand\prd{Phys.~Rev.~D}
\newcommand\sovast{Soviet~Ast.}
\newcommand\nat{Nature}
\newcommand\jcap{Journal of Cosmology and Astroparticle Physics}

\title[Peaks and dips in Gaussian random fields]{Peaks and dips in Gaussian random fields: a new algorithm  for the shear eigenvalues, and the excursion set theory}
\author[Graziano Rossi] {Graziano Rossi$^{1,2,3}$\thanks{E-mail: graziano.rossi@cea.fr} \\
\\
$^{1}$ Korea Institute for Advanced Study, Hoegiro 85, Dongdaemun-Gu, Seoul $130-722$, Korea \\
$^{2}$ CEA, Centre de Saclay, Irfu/SPP, F-91191 Gif-sur-Yvette, France \\
$^{3}$ Paris Center for Cosmological Physics (PCCP) and Laboratoire APC, Universit\'e Paris 7, 75205 Paris, France}
\date{Accepted 2012 November 27. Received 2012 November 7; in original form 2012 June 29}
\pagerange{\pageref{firstpage}--\pageref{lastpage}}
\pubyear{2013}

\begin{document}
\maketitle
\label{firstpage}



\begin{abstract}

We present a new algorithm to sample the constrained eigenvalues of the initial shear field associated with Gaussian statistics, called 
the `peak/dip excursion-set-based' algorithm, at positions which
correspond to peaks or dips of the correlated density field.  
The computational procedure is based on a new formula which extends Doroshkevich's unconditional distribution for the eigenvalues of the linear tidal field,
 to account for the fact that halos and voids may correspond to 
maxima or minima of the density field.
The ability to differentiate between random positions and special points in space around which halos or voids may form (i.e. peaks/dips),
encoded in the new formula and reflected in the algorithm, naturally leads 
to a straightforward implementation of an excursion set model for peaks and dips in Gaussian random fields -- one of the key advantages of this sampling procedure.
In addition, it offers novel insights into the statistical description of the cosmic web. 
As a first physical application, we show how the standard distributions of shear ellipticity and prolateness 
in triaxial models of structure formation
are modified by the constraint. In particular, we 
provide a new expression for the 
conditional distribution of shape parameters given the density peak constraint,  which generalizes some previous literature work.
The formula has important implications for the modeling of non-spherical dark matter halo shapes, in relation to their initial shape distribution.  
We also test and confirm our theoretical predictions for the individual distributions of eigenvalues
subjected to the extremum constraint, along with other directly related conditional probabilities.
Finally, we indicate how the proposed sampling procedure naturally integrates into the standard excursion set model, potentially solving some of its well-known problems, and into the
ellipsoidal collapse framework.
Several other ongoing applications and extensions, towards the development of 
algorithms for the morphology and topology of the cosmic web, are discussed at the end. 

\end{abstract}






\section{Introduction}


The large-scale spatial organization of matter in clusters, filaments, sheets and voids, known as the
\textit{cosmic web} (Peebles 1980;  Bardeen et al. 1986;  Bond et al. 1991), is the manifestation
of the anisotropic nature of gravitational collapse.  This typical filamentary pattern has been confirmed by 
observations, for example with the
2dFGRS, SDSS and 2MASS redshift surveys of the nearby Universe 
(Colless et al. 2003; Tegmark et al. 2004; Huchra et al. 2005), and is routinely seen in large scale $N$-body 
numerical simulations -- see for example the  New Horizon Runs (Kim et al. 2011) or the recent Millennium-XXL (Angulo et al. 2012).
Basic characteristics of the cosmic web are the 
spatial arrangement of galaxies and mass into elongated filaments, sheet-like walls and dense compact clusters,
the existence of large near-empty void
regions, and the hierarchical nature of this mass distribution (Arag{\'o}n-Calvo et al. 2007, 2010a,b; Zhang et al. 2009). In particular, 
as pointed out by Bond, Kofman \& Pogosyan (1996),  `embryonic' cosmic web is already present in the primordial density field.
These key properties, along with
the alignment of shape and angular momentum of objects 
(i.e. Argyres et al. 1986; Catelan et al. 2001; Lee  \& Springel 2010),
are mainly due to the effects of the tidal field, associated with the gravitational potential -- while the 
Hessian of the density field (i.e. the inertia tensor) plays a secondary role in determining the characteristic pattern of the cosmic web. 


Pioneering works 
devoted to the key role of the initial tidal field in shaping large-scale structures trace back to Doroshkevich \&
Zeldovich (1964),  Doroshkevich (1970), Zeldovich (1970),  Sunyaev \& Zeldovich (1972),  Icke (1973), and Doroshkevich \& Shandarin (1978); their studies
have contributed to reach a solid understanding of the formation and evolution of structures.  
Other classical works in this direction (i.e. Peebles 1980;   
White 1984; Bardeen et al. 1986; Kaiser 1986;  Bertschinger 1987;  Dubinski 1992; Bond \& Myers 1996; Bond, Kofman \& Pogosyan 1996;  van de Weygaert \& Bertschinger 1996) 
have considerably improved the statistical description of the cosmic web from first principles. Their impact  is broader, as
 there is a correspondence between structures in the evolved density field and local properties
of the linear tidal shear; this allows one to estimate the morphology of the cosmic web (Bond \& Myers 1996;  Rossi, Sheth \& Tormen 2011), and is crucial in understanding the
nonlinear evolution of cosmic structures
(Springel et al. 2005; Shandarin et al. 2006; Pogosyan et al. 2009), 
the statistical properties of voids (Lee \&
Park 2006; Platen, van de Weygaert \& Jones 2008; Lavaux \& Wandelt 2010),
the alignment of shape and angular
momentum of halos (West 1989; Catelan et al. 2001; Faltenbacher et al. 2009; Lee \& Springel 2010), 
and for characterizing the geometry and topology of the cosmic web
(Gott et al. 1986, 1989; Park \& Gott 1991; Park et al. 1992, 2005;  van de Weygaert \& Bond 2008; Forero-Romero et al. 2009; Aragon-Calvo et al. 2010a,b; Choi et al. 2010; Shandarin et al. 2010; van de Weygaert et al. 2011; Cautun et al. 2012; Hidding et al. 2012). 
In addition, the eigenvalues of the mass tidal tensor can be used to classify the large-scale environment  (Shen et al. 2006; Hahn et al. 2007; Zhang et al. 2009) -- hence their fundamental importance.


In the context of the initial shear field, 
Doroshkevich (1970) provided a key contribution by deriving  
 the joint probability distribution of an ordered set of eigenvalues in the tidal
field matrix corresponding to a Gaussian potential, given the variance of the density field. 
Throughout the paper, we will refer to it as the \textit{unconditional} probability distribution of shear eigenvalues. 
Because the initial shear field associated with Gaussian statistics plays a major role in the formation of
large scale structures, considerable analytic work has been based 
on  Doroshkevich's unconditional formulas since their appearance, but
those relations neglect the fact that 
halos (voids) may correspond to
maxima (minima) of the underlying density field.  
The local extrema of such field (i.e. peaks/dips) are plausible sites for the formation of
nonlinear structures (Bardeen et al. 1986), and some numerical studies 
have indeed reported a good correspondence 
between peaks in the initial conditions and halos at late times (see in particular Ludlow \& Porciani 2011). 
Hence, their statistical properties can be
used to predict the abundances and clustering of objects of
various types, and in studies of triaxial formation of
large-scale structures (Bardeen et al. 1986; Bond \& Myers 1996).


Motivated by these reasons, recently Rossi (2012)
has provided a set of analytic expressions which extend
the work of Doroshkevich (1970) and Bardeen et al. (1986) -- and are akin in philosophy to that of
van de Weygaert \& Bertschinger (1996). 
These new relations 
incorporate the 
peak/dip constraint into the statistical description of the 
initial shear field, and are able to differentiate between random positions and peak/dips in the correlated 
density field.
In essence, they allow one to express the 
joint probability distribution of an ordered set of eigenvalues
in the initial shear field, given the fact that positions are peaks
or dips of the density -- and not just random spatial
locations.
These relations are
obtained by 
requiring the density Hessian (i.e. the matrix of the second derivatives of the density field, associated with the curvature) to be positive/negative definite, which is
the case in the vicinity of minima/maxima of the density. 
The correlation strength between the gravitational and
density fields is quantified via
a reduced parameter $\gamma$, which plays a major role in peaks theory  (i.e. $\gamma$ is the same as in Bardeen et al. 1986 for Gaussian smoothing filters 
-- see Appendix \ref{notation}). Doroshkevich's (1970)
\textit{unconditional} formulas are then naturally recovered in the absence of correlation, when $\gamma=0$. 
From these new \textit{conditional} joint probabilities, it is possible to derive the individual distributions of eigenvalues subjected to
the extremum constraint, along with some other related conditional probabilities; their expressions were also
provided in Rossi (2012), extending the work of Lee \& Shandarin (1998). 


The primary goal of this paper is show how the main conditional formulas presented 
in Rossi (2012) and reported here (Equations \ref{doro_inter_extended} and \ref{doro_eigen}) lead to  a  
new algorithm -- called the `peak/dip excursion-set-based' algorithm -- to sample the constrained eigenvalues of the initial shear field associated with Gaussian statistics,  
at positions which correspond to peaks or dips of the correlated density field. While it is clearly 
possible to sample the constrained eigenvalues of the 
tidal field directly from the conditional probability distribution function, as done for example by Lavaux \&  Wandelt (2010) in the context of cosmic voids, 
the theoretical work carried out by Rossi (2012) allows for a 
much simpler procedure,  part of which was previously thought not to be achievable analytically (see again the Appendix B in Lavaux \&  Wandelt 2010). 
Besides providing novel theoretical insights, the main strength of the
algorithm resides in its natural inclusion into the standard excursion set framework, allowing for a generalization.  
As we will discuss in Section \ref{esmpd},  this technique potentially solves  a long-standing problem of the standard excursion set theory; moreover, 
it is well-suited for the ellipsoidal collapse framework.  
 
The other goal of the paper is to present a first physical application of the new algorithm, related to the morphology of the cosmic web.
In particular, we provide a new expression for the 
conditional distribution of shape parameters (i.e. ellipticity and prolateness) in the presence of the density peak constraint, 
which generalizes some previous literature work and combines
the formalism of Bardeen et al. (1986) -- based on the density field  -- with that of Bond \& Myers (1996) -- based on the shear field.
The formula has important implications for the modeling of non-spherical dark matter halos and their evolved halo shapes, in relation to the initial shape distribution.  
In addition, we also test and confirm our theoretical predictions for the individual distributions of eigenvalues 
subjected to the extremum constraint, along with other directly related conditional probabilities.
Finally, we illustrate how this algorithm can be readily merged into the excursion set framework (Peacock \& Heavens 1990; Bond et al. 1991), and in particular to obtain 
an excursion set model for peaks and dips in Gaussian random fields. 
The key point is the 
 ability to differentiate between random positions and peaks/dips, which is contained in Equations  (\ref{doro_inter_extended}) and (\ref{doro_eigen}) and encapsulated in the algorithm.
While we leave this latter part to a dedicated forthcoming publication, we anticipate the main ideas here. We also discuss
several other ongoing applications and extensions, towards the development of 
algorithms for classifying cosmic web structures. 


The layout of the paper is organized as follows. 
Section \ref{jde} provides a short review of the key equations derived in Rossi (2012), which constitute the theoretical framework for the new algorithm described here;
the main notation adopted is summarized in Appendix \ref{notation}, for convenience.
Moving from this mathematical background, Section \ref{pdesa}  presents the new `peak/dip excursion-set-based' algorithm (some insights derived from this part
are left in Appendix \ref{insights}), while Section 
\ref{cdpnt} tests its performance against various analytic distributions of eigenvalues and related conditional probabilities,
subjected to the extremum constraint. 
Section \ref{epd} shows a physical application of the computational procedure, towards the morphology of the cosmic web.
A new expression for the conditional distribution of ellipticity and prolateness in the presence of the density peak constraint is also given; in particular,
the description of Bardeen et al. (1986) is combined with that of Bond \& Myers (1996).
Section \ref{esmpd} illustrates how this new algorithm can be readily used to implement an excursion set model for peaks and dips in Gaussian random fields, and makes the connection with
some previous literature. 
Finally, Section \ref{conc} discusses several ongoing and future promising applications, which will be presented in forthcoming publications, and in particular
 the use of this algorithm for triaxial models of collapse and in relation to the morphology and topology of the cosmic web.



\section{Joint conditional distribution of eigenvalues in the peak/dip picture}
\label{jde}


We begin by reexamining two main results derived in Rossi (2012), which constitute
the key for developing a new algorithm to sample the constrained eigenvalues of the initial shear field. This part may be also regarded as a compact review of the main formulas for
the constrained shear eigenvalues, which can be used directly for several applications related to the cosmic web.
The notation adopted here is the same as the one introduced by Rossi (2012),  with a few minor changes to make the
connection with previous literature more explicit. It is  summarized in Appendix \ref{notation}; a reader not familiar with the notation
may want to start from the appendix first. In particular, in what follows
 we do not adopt `reduced' variables, so that the various dependencies on $\sigma$ values (i.e. $\sigma_{\rm T} \equiv \sigma_0$ for the
 shear, and $\sigma_{\rm H} \equiv \sigma_2$ for the density Hessian) are now shown explicitly.
 However, for the sake of clarity, we omit to indicate the understood dependence of these two global parameters 
 in the left-hand side of all the formulas.
  Note that we prefer to write $\sigma_{\rm T}$ and $\sigma_{\rm H}$, rather than $\sigma_0$ and $\sigma_2$, for their more intuitive meaning 
  (i.e. the label $T$ indicates that a quantity is connected to the shear field, while the label $H$ points to the Hessian of the density field).
As explained in Appendix  \ref{notation}, we also introduce the 6-dimensional vectors ${\bf T} = (T_{11},T_{22},T_{33},T_{12},T_{13},T_{23})$ and ${\bf H} = (H_{11},H_{22},H_{33},H_{12},H_{13},H_{23})$,
derived from the components of their corresponding shear and density Hessian tensors.
  
Rossi (2012) obtained the following expression for the probability of observing a tidal field ${\bf T}$ for the gravitational potential, given a curvature
${\bf H}$ for the density field:
\begin{equation}
p ({\bf T}|{\bf H},\gamma) = {15^3 \over 16 \sqrt {5} \pi^3} {1 \over \sigma^6_{\rm T}(1-\gamma^2)^3} {\rm exp} \Big [-{3 \over 2 \sigma_{\rm T}^2(1-\gamma^2)} (2 K_1^2 - 5 K_2 ) \Big ]
\label{doro_inter_extended}
\end{equation}
where 
\begin{eqnarray}
\label{K_def}
K_1 &=& (T_{11} -\eta H_{11}) + (T_{22} -\eta H_{22}) + (T_{33} -\eta H_{33}) = k_1 - \eta h_1 \\
K_2 &=& (T_{11} -\eta H_{11})(T_{22} - \eta H_{22})  + (T_{11} -\eta H_{11})(T_{33} - \eta H_{33}) \nonumber \\
&+& (T_{22} -\eta H_{22})(T_{33} - \eta H_{33}) -(T_{12} - \eta H_{12})^2 
-(T_{13} -\eta H_{13})^2 -(T_{23} - \eta
H_{23})^2 = k_2 + \eta^2 h_2 - \eta h_1 k_1 + \eta \tau \\
\tau &=& T_{11} H_{11} + T_{22} H_{22} + T_{33} H_{33} + 2 T_{12} H_{12} + 
2 T_{13} H_{13} + 2 T_{23} H_{23} \\
k_1 &=& T_{11} + T_{22} + T_{33} \\
k_2 &=&  T_{11} T_{22}+  T_{11}T_{33}+ T_{22}T_{33} -T^2_{12} -T^2_{13} -T^2_{23}\\ 
h_1 &=& H_{11} + H_{22} + H_{33} \\
h_2 &=&  H_{11} H_{22}+ H_{11} H_{33}+ H_{22} H_{33} -H^2_{12} -H^2_{13} -H^2_{23}\\
\eta &=& \gamma \sigma_{\rm T} / \sigma_{\rm H}.
\end{eqnarray}
The corresponding unconditional
marginal distributions $p({\bf T})$ and $p({\bf H})$ are multidimensional Gaussians, expressed using Doroshkevich's formulas as
\begin{equation}
p({\bf T}) = {15^3 \over 16 \sqrt{5} \pi^3} {1 \over \sigma^6_{\rm T} } {\rm exp} \Big [ {- \rm {3 \over 2 \sigma^2_{\rm T}} (2
  k_1^2 - 5 k_2)} \Big ],~~ p({\bf H}) = {15^3 \over 16 \sqrt{5} \pi^3} {1 \over \sigma_{\rm H}^6} {\rm exp} \Big [{- \rm {3 \over 2 \sigma^2_{\rm H}} (2 h_1^2 - 5 h_2)} \Big ].
\label{doro_standard}
\end{equation}
Equation (\ref{doro_inter_extended})  generalizes
Doroshkevich's formulas (\ref{doro_standard}) to include 
the fact that halos/voids may correspond to
maxima/minima of the density field. 
Note in particular that  $p({\bf T}|{\bf H}, \gamma)$ is a multivariate Gaussian distribution with mean $b = \eta {\bf H}$ and covariance matrix
$(1-\gamma^2) \sigma^2_{\rm T} {\bf \sf A}/15$, with ${\bf \sf A}$ given in Appendix \ref{notation}.
Clearly, one can also consider the reverse distribution $p ({\bf H}|{\bf T}, \gamma)$,
the expression of which is given in Rossi (2012). This is useful in order to make the connection and extend some results derived in Bardeen et al. (1986), 
the subject of a forthcoming publication.

It is also possible to express (\ref{doro_inter_extended}) in terms of
the constrained eigenvalues of ${\bf T}|{\bf H}$, indicated with $\zeta_{\rm i}$ ($i=1,2,3$) and ordered as $\zeta_1 \ge \zeta_2 \ge \zeta_3$.
The result is:
\begin{equation}
\label{doro_eigen}
p(\zeta_1,
  \zeta_2, \zeta_3|\gamma)
= {15^3 \over 8 \sqrt {5} \pi} {1 \over \sigma^6_{\rm T} (1-\gamma^2)^3}
 {\rm exp} \Big [-{3 \over 2 \sigma^2_{\rm T} (1-\gamma^2)} (2 K_1^2 - 5 K_2 ) \Big ] 
(\zeta_1-\zeta_2) (\zeta_1-\zeta_3) (\zeta_2-\zeta_3)
\end{equation}
where in terms of constrained eigenvalues we now have:
\begin{eqnarray}
\label{eigen_doro_ext}
K_1 &=& \zeta_1 + \zeta_2 +\zeta_3 =   k_1 - \eta h_1 \\
K_2 &=&  \zeta_1 \zeta_2 + \zeta_1 \zeta_3 + \zeta_2 \zeta_3 
 = k_2 + \eta^2 h_2 - \eta h_1 k_1 + \gamma \tau \\
\tau &=& \lambda_1 \xi_1 +  \lambda_2 \xi_2 +  \lambda_3 \xi_3 \\  
k_1 &=&  \lambda_1 + \lambda_2 + \lambda_3 \\
k_2 &=&  \lambda_1 \lambda_2 +  \lambda_1 \lambda_3 +  \lambda_2 \lambda_3 \\
h_1 &=&  \xi_1 + \xi_2 + \xi_3 \\
h_2 &=&  \xi_1 \xi_2 +\xi_1 \xi_3 + \xi_2 \xi_3\\
\zeta_{\rm i} &=& \lambda_{\rm i} - \eta \xi_{\rm i}.
\end{eqnarray}
The partial distributions $p(\lambda_1,\lambda_2,\lambda_3)$ and  $p(\xi_1,\xi_2,\xi_3)$
are expressed by Doroshkevich's
unconditional formulas as
\begin{equation}
p(\lambda_1,\lambda_2,\lambda_3) = {15^3 \over
    8 \sqrt{5} \pi}{1 \over \sigma^6_{\rm T}}  {\rm exp} \Big [{\rm - {3
    \over 2 \sigma^2_{\rm T}} (2 k_1^2 - 5 k_2)}  \Big ] (\lambda_1-\lambda_2)
    (\lambda_1-\lambda_3) (\lambda_2-\lambda_3)
\label{doro_original}
\end{equation}
\begin{equation}
p(\xi_1,\xi_2,\xi_3) = {15^3 \over 8 \sqrt{5}
    \pi} {1 \over \sigma^6_{\rm H}}  {\rm exp} \Big [{\rm - {3 \over 2 \sigma^2_{\rm H}} (2 h_1^2 - 5 h_2)} \Big ]  (\xi_1-\xi_2)
    (\xi_1-\xi_3) (\xi_2-\xi_3),
    \label{doro_original_bis}
\end{equation}
where $\lambda_1, \lambda_2, \lambda_3$ are the eigenvalues of the shear tensor, while 
$\xi_1, \xi_2, \xi_3$ are those of the density Hessian.
Similarly, one can easily obtain formulas for the reverse probability functions. Note also that $k_1=\lambda_1+ \lambda_2+\lambda_3$ is simply the overdensity $\delta_{\rm T}$ 
associated to the shear field,
while $\xi_1 + \xi_2 + \xi_3 = h_1 \equiv \delta_{\rm H}$. Later on, we will make use of the
peak height $\nu = \delta_{\rm T}/ \sigma_{\rm T}$ and of the peak curvature  $x = \delta_{\rm H}/ \sigma_{\rm H}$. 

Equations (\ref{doro_inter_extended}) and (\ref{doro_eigen}) allow one to develop
a new algorithm to sample the constrained eigenvalues of the initial shear field, presented in the next section. 
It will be then straightforward to use this algorithm in order to implement
an excursion set model for peaks and dips in Gaussian random fields.
 


\section{The peak/dip excursion-set-based algorithm}
\label{pdesa}


While the constrained eigenvalues of the initial shear field can be sampled directly from their probability distribution function
(i.e. Equations \ref{doro_inter_extended} or \ref{doro_eigen}), the new theoretical formalism developed by Rossi (2012) leads to a much simpler
algorithm, which is particularly interesting because well-suited for the ellipsoidal collapse model (Section \ref{epd}), and naturally integrable in
the excursion set framework (Section \ref{esmpd}). 
In addition, the algebra leading to the new computational procedure allows one to explain analytically how the halo (void) shape distributions
are altered by the inclusion of the peak (dip) constraint, and offers a variety of applications and insights on the statistical description of the cosmic web (see Section \ref{conc} and Appendix \ref{insights}).
In what follows,  we present the mathematical aspects of the `peak/dip excursion-set-based' algorithm, and describe in detail the novel computational procedure.

In particular, we are mainly interested in the distribution $p({\bf T}|{\bf H}>0,\gamma)$, the joint probability 
of observing a tidal field ${\bf T}$ for the gravitational potential with a positive density curvature ${\bf H}>0$ (i.e. at density peak locations). 
Clearly,
\begin{equation}
\label{peak_dist}
p({\bf T}| {\bf H}>0, \gamma) = { p({\bf T}, {\bf H}>0| \gamma)  \over p({\bf H}>0)}  =
{ \int_{{\bf H}>0}  p({\bf H}) ~ p({\bf T}|{\bf H}, \gamma) ~{\rm d}{\bf H} \over  \int_{{\bf H}>0}  p({\bf H})~{\rm d}{\bf H}},
\end{equation}
where $p({\bf T}|{\bf H}, \gamma)$ and  $p({\bf H})$ are given by Equations (\ref{doro_inter_extended}) and (\ref{doro_standard}), and the integrals are 6-dimensional.
A similar expression can be written in terms of the constrained distributions of eigenvalues, using Equation (\ref{doro_eigen}) instead.
In principle, one should then compute the previous integral to obtain $p({\bf T}| {\bf H}>0, \gamma)$. However, 
 it is easier to sample $p({\bf T}|{\bf H}, \gamma)$ and impose the condition ${\bf H}>0$ directly, so that we are effectively computing
$p({\bf T}|{\bf H}>0, \gamma)$.  Given the nature of $p({\bf T}|{\bf H}, \gamma)$ and $p({\bf H})$
expressed by (\ref{doro_inter_extended}) and (\ref{doro_standard}),  
this can be readily achieved -- 
as the following algebra will show.  

In fact, because  the elements of the density Hessian are drawn from a multivariate
Gaussian distribution with zero mean and covariance matrix $\sigma^2_{\rm H}{\bf \sf A}/15$,
where ${\bf \sf A}$ is given by Equation (\ref{matrix_A}),
one can simply obtain them by generating six independent zero-mean unit-variance Gaussian random variates $y_{\rm i}$ ($i=1,6$),  and then determine the 
various components as:
\begin{eqnarray}
\label{H_gauss}
H_{11}&= & {\sigma_{\rm H} \over 3}   \Big ( -y_1 +{2 \over \sqrt{5}} y_2 \Big )  \nonumber \\
H_{22}&= &    {\sigma_{\rm H}  \over 3}    \Big ( - y_1 - {1 \over \sqrt{5}} y_2 - {3 \over \sqrt{15} } y_3  \Big )     \nonumber \\
H_{33}&= &    {\sigma_{\rm H}  \over 3}   \Big ( - y_1 - {1 \over \sqrt{5}} y_2 + {3 \over \sqrt{15} } y_3  \Big )     \nonumber \\
H_{12}& = &  H_{21} =  {\sigma_{\rm H}  \over \sqrt{15}} y_4  \nonumber \\
H_{13}&= &   H_{31}  =   {\sigma_{\rm H}  \over \sqrt{15}} y_5  \nonumber \\
H_{23}&= &   H_{32}  = {\sigma_{\rm H}  \over \sqrt{15}}  y_6. 
\end{eqnarray}

Similarly, the elements of the shear tensor are also drawn from a
multivariate Gaussian distribution with mean zero and covariance matrix
$\sigma^2_{\rm T}{\bf \sf A}/15$. Hence, if 
$z_{\rm i}$ ($i=1,6$) are other six independent zero-mean unit-variance Gaussian random variates, the
components of the shear tensor are given by:
\begin{eqnarray}
\label{T_gauss}
T_{11}&= & {\sigma_{\rm T} \over 3}   \Big ( -z_1 +{2 \over \sqrt{5}} z_2 \Big )  \nonumber \\
T_{22}&= &    {\sigma_{\rm T}  \over 3}    \Big ( - z_1 - {1 \over \sqrt{5}} z_2 - {3 \over \sqrt{15} } z_3  \Big )     \nonumber \\
T_{33}&= &    {\sigma_{\rm T}  \over 3}   \Big ( - z_1 - {1 \over \sqrt{5}} z_2 + {3 \over \sqrt{15} } z_3  \Big )     \nonumber \\
T_{12}&= & T_{21} =  {\sigma_{\rm T}  \over \sqrt{15}} z_4  \nonumber \\
T_{13}&= & T_{31} =  {\sigma_{\rm T}  \over \sqrt{15}} z_5  \nonumber \\
T_{23}&= & T_{32} = {\sigma_{\rm T}  \over \sqrt{15}}  z_6. 
\end{eqnarray}

Next, consider the 6-dimensional vector  ${\bf T}|{\bf H}$, made from the components of the conditional shear field. Its elements are drawn from a
multivariate Gaussian distribution $p({\bf T}|{\bf H}, \gamma)$ 
with mean $b = \eta {\bf H}$ and covariance matrix
$(1-\gamma^2) \sigma^2_{\rm T} {\bf \sf A}/15$, where the elements of the density Hessian are expressed by (\ref{H_gauss}). 
Therefore, this implies that the
  components  of ${\bf T}|{\bf H}$ are distributed according to:
\begin{eqnarray}
\label{T_given_H_gauss}
T_{11}|H_{11} &= &     \eta H_{11} + {\sigma_{\rm T}\sqrt{1 - \gamma^2} \over 3}   \Big ( -l_1 +{2 \over \sqrt{5}} l_2 \Big ) \equiv 
 {\sigma_{\rm T} \over 3}   \Big ( -m_1 +{2 \over \sqrt{5}} m_2 \Big ) \nonumber \\
T_{22}|H_{22}   &= &     \eta H_{22} +{\sigma_{\rm T} \sqrt{1-\gamma^2} \over 3}    \Big ( - l_1 - {1 \over \sqrt{5}} l_2 - {3 \over \sqrt{15} } l_3  \Big ) \equiv {\sigma_{\rm T} \over 3}    \Big ( - m_1 - {1 \over \sqrt{5}} m_2 - {3 \over \sqrt{15} } m_3  \Big )       \nonumber \\
 T_{33}|H_{33}  &= &    \eta H_{33} + {\sigma_{\rm T} \sqrt{1-\gamma^2} \over 3}   \Big ( - l_1 - {1 \over \sqrt{5}} l_2 + {3 \over \sqrt{15} } l_3  \Big )    \equiv  {\sigma_{\rm T} \over 3}   \Big ( - m_1 - {1 \over \sqrt{5}} m_2 + {3 \over \sqrt{15} } m_3  \Big )   \nonumber \\
 T_{12}|H_{12} &= &   T_{21}|H_{21} =  \eta H_{12} +{\sigma_{\rm T} \sqrt{1-\gamma^2} \over \sqrt{15}} l_4  \equiv  {\sigma_{\rm T} \over \sqrt{15}} m_4  \nonumber \\
 T_{13}|H_{13} &= & T_{31}|H_{31} =  \eta H_{13} +{\sigma_{\rm T} \sqrt{1-\gamma^2} \over \sqrt{15}} l_5   \equiv  {\sigma_{\rm T} \over \sqrt{15}} m_5 \nonumber \\
 T_{23}|H_{23} &= &  T_{32}|H_{32}  =  \eta H_{23} + {\sigma_{\rm T} \sqrt{1-\gamma^2} \over \sqrt{15}}  l_6  \equiv  {\sigma_{\rm T} \over \sqrt{15}} m_6.
\end{eqnarray}
In the previous expression, the various $l_{\rm i}$ ($i=1,6$) are  other six independent zero-mean unit-variance Gaussian variates,
while the $m_{\rm i}$ ($i=1,6$) are six independent Gaussian distributed variates with shifted mean $\gamma y_{\rm i}$ and reduced variance $(1-\gamma^2)$, i.e. $m_{\rm  i} = \gamma y_{\rm i}  + \sqrt{1-\gamma^2} ~ l_{\rm i} $.

Equations  (\ref{H_gauss}), (\ref{T_gauss}) and  (\ref{T_given_H_gauss}) 
suggest a new excursion-set-based
algorithm,  in order to obtain the constrained eigenvalues of the matrix having components $T_{\rm \alpha}|H_{\rm \alpha}$ 
(see  Appendix \ref{notation} for the definition of the index $\alpha$)  -- supplemented by 
the condition of a positive curvature ${\bf H}$ for the density field  (or negative curvature, for voids). 
The procedure can be summarized as follows: 

\begin{enumerate}

\item Draw six independent zero-mean unit-variance  Gaussian  distributed
variates $y_{\rm i}$  ($i=1,6$), and 
determine the components $H_{\rm \alpha}$ of the density Hessian matrix via Equation (\ref{H_gauss}). Compute the value of $\sigma_{\rm H}$ using (\ref{eq_sig}).
\item  Calculate the eigenvalues $\xi_1, \xi_2, \xi_3$ of the previous Hessian matrix, and check if they are all positive (negative). If so, proceed to the next step, otherwise try again.
This will guarantee the Hessian to be positive (negative) definite (i.e. the Hessian is a real-symmetric matrix), which is the condition for maxima (minima) of the field. Note that this step 
is clearly not required if we want to sample only $p({\bf T}|{\bf H}, \gamma)$.
\item Draw other six independent Gaussian  distributed
variates  $l_{\rm i}$ ($i=1,6$), with mean zero and variance one, and determine the $T_{\rm \alpha}|H_{\rm \alpha}$ components via  Equation (\ref{T_given_H_gauss}), using the 
previously accepted
values $H_{\rm \alpha}$ from (\ref{H_gauss}) -- while $\sigma_{\rm T}$ is determined via (\ref{eq_sig}). 
Since we require the density Hessian to be positive definite, this means that we are effectively
sampling the conditional probability $p({\bf T}|{\bf H}>0, \gamma)$ -- or $p({\bf T}|{\bf H}<0, \gamma)$ for a negative definite Hessian.
\item Calculate and store  the constrained eigenvalues $\zeta_{\rm i}$ of the matrix having components $T_{\rm \alpha}|H_{\rm \alpha}>0$.

\end{enumerate}

We leave in Appendix  \ref{insights}
some more insights on the main conditional formula (\ref{doro_inter_extended}), which readily follow from  the 
previous Equations  (\ref{H_gauss}), (\ref{T_gauss}) and  (\ref{T_given_H_gauss}). In the next sections, we 
will show a few applications of the new algorithm -- with particular emphasis on the 
conditional distributions of shape parameters in triaxial models of structure formation (i.e. ellipticity and prolateness).
Additional applications will be presented in forthcoming publications. 



\section{Conditional distributions and probabilities: numerical tests}
\label{cdpnt}


The algorithm illustrated in the previous section allows one
to test and confirm several formulas derived by Rossi (2012), regarding conditional distributions and probabilities subjected to the extremum constraint. 
We present here results of the comparison (i.e. theory versus numerical), 
where for simplicity  we set $\sigma_{\rm T} = \sigma_{\rm H} \equiv 1$;
this corresponds to adopt `reduced variables', as done in Rossi (2012). Note that we show explicitly the 
dependencies on $\sigma$ values in the following expressions, although we set them to unity in the mock tests.
 

\subsection{Individual conditional distributions}

The individual conditional distributions of
shear eigenvalues,
for a given
correlation strength $\gamma$ with the
density field (which encapsulate the peak/dip constraint), 
are given by (Rossi 2012): 
\begin{eqnarray}
\label{zeta1}
p(\zeta_1|\gamma) 
 &=& {\sqrt{5} \over 12 \pi \sigma_{\rm T}} \Big [ {20 \over (1-\gamma^2)}  {\zeta_1 \over \sigma_{\rm T}} ~{\rm exp} \Big (-
  {9 \over 2 (1-\gamma^2)   } {\zeta_1^2 \over \sigma^2_{\rm T}  } 
  \Big ) - {\sqrt{2 \pi} \over (1-\gamma^2)^{3/2}}~{\rm exp} \Big(- {5 
    \over 2 (1-\gamma^2)} {\zeta_1^2 \over \sigma^2_{\rm T}}  \Big
  ) \times \\
 && {\rm erfc} \Big (-{\sqrt{2}  \over \sqrt{1-\gamma^2}} {\zeta_1 \over \sigma_{\rm T}}
  \Big ) \Big [ (1-\gamma^2)-20 {\zeta_1^2 \over \sigma^2_{\rm T}} \Big ] 
+ {3 \sqrt{3 \pi} \over \sqrt{1-\gamma^2}}~
       {\rm exp} \Big (- {15  \over 4 (1-\gamma^2)}{\zeta^2_1 \over \sigma^2_{\rm T}}  \Big ) {\rm erfc} \Big (
       -{\sqrt{3} \over 2 \sqrt{1-\gamma^2}}{\zeta_1 \over \sigma_{\rm T}} 
       \Big )        
             \Big ] \nonumber
 \end{eqnarray}

\begin{eqnarray}
\label{zeta2}
p(\zeta_2|\gamma) &=& {\sqrt{15} \over 2 \sqrt{\pi}} {1 \over \sigma_{\rm T}\sqrt{1-\gamma^2}}
{\rm exp} \Big [- {15 \over 4 (1 -\gamma^2)}
  {\zeta_2^2 \over \sigma^2_{\rm T} }  \Big ] 
\end{eqnarray}

\begin{eqnarray}
\label{zeta3}
p(\zeta_3|\gamma) 
 &=& -{\sqrt{5} \over 12 \pi \sigma_{\rm T}} \Big [ {20 \over (1-\gamma^2)}  {\zeta_3 \over \sigma_{\rm T}} ~{\rm exp} \Big (-
  {9 \over 2 (1-\gamma^2)   } {\zeta_3^2 \over \sigma^2_{\rm T}  } 
  \Big ) + {\sqrt{2 \pi} \over (1-\gamma^2)^{3/2}}~{\rm exp} \Big(- {5 
    \over 2 (1-\gamma^2)} {\zeta_3^2 \over \sigma^2_{\rm T}}  \Big
  ) \times  \\
 && {\rm erfc} \Big ({\sqrt{2}  \over \sqrt{1-\gamma^2}} {\zeta_3 \over \sigma_{\rm T}}
  \Big ) \Big [(1-\gamma^2)-20 {\zeta_3^2 \over \sigma^2_{\rm T}} \Big ] 
- {3 \sqrt{3 \pi} \over \sqrt{1-\gamma^2}}~
       {\rm exp} \Big (- {15  \over 4 (1-\gamma^2)}{\zeta^2_3 \over \sigma^2_{\rm T}}  \Big ) {\rm erfc} \Big (
       {\sqrt{3} \over 2 \sqrt{1-\gamma^2}}{\zeta_3 \over \sigma_{\rm T}} 
       \Big )        
             \Big ]. \nonumber
 \end{eqnarray}

\begin{figure}
\centering
\includegraphics[angle=0,width=1.00\textwidth]{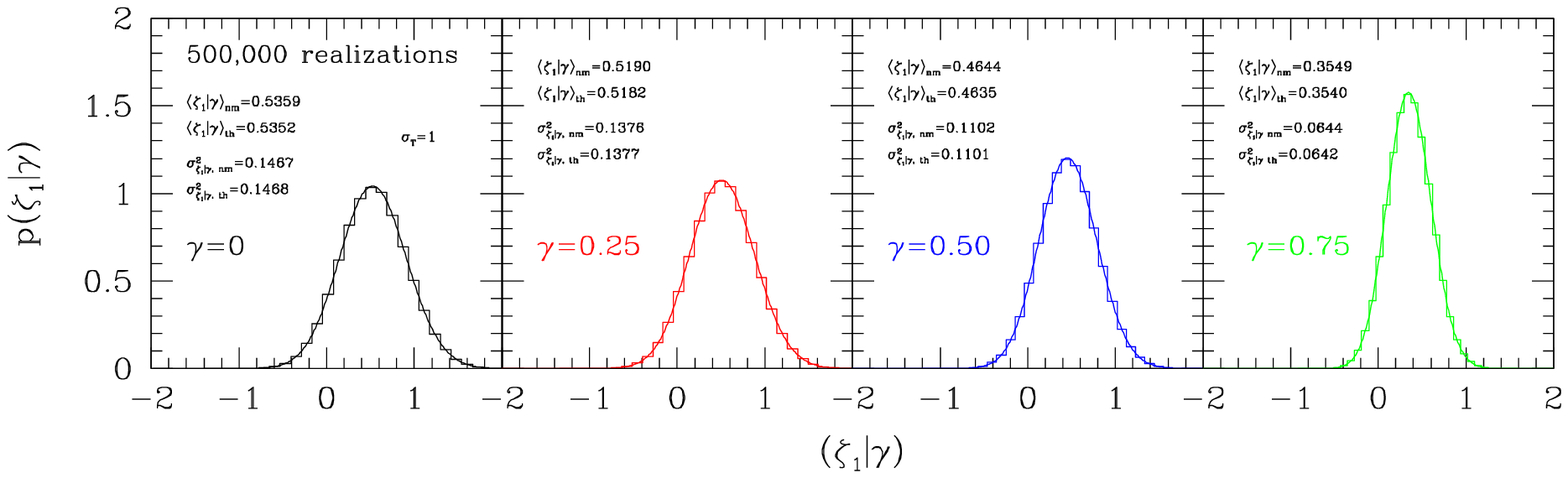}
\includegraphics[angle=0,width=1.00\textwidth]{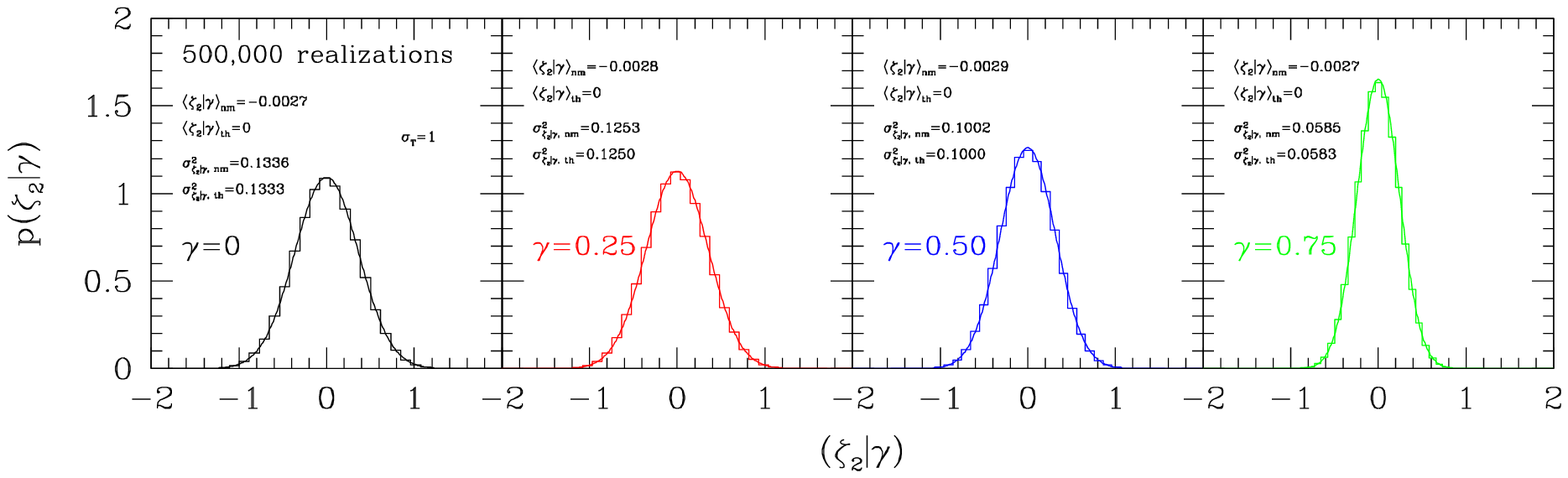}
\includegraphics[angle=0,width=1.00\textwidth]{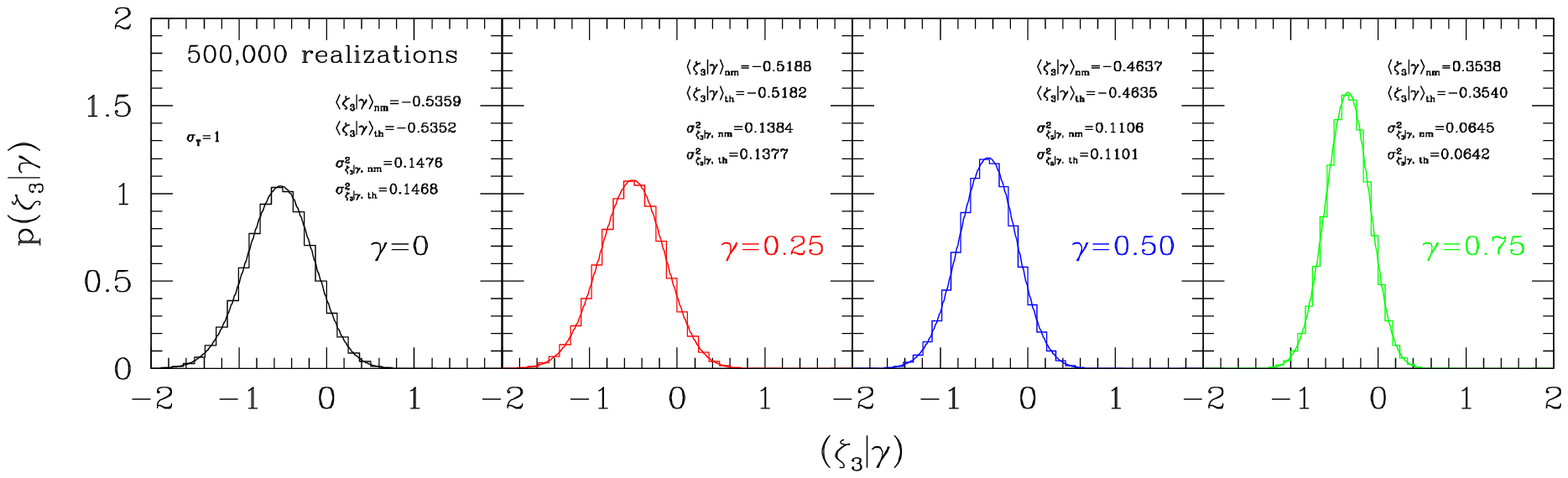}
\caption{Individual conditional probabilities $p(\zeta_1|\gamma)$ [top],  $p(\zeta_2|\gamma)$ [middle], and
$p(\zeta_3|\gamma)$ [bottom] of the initial shear field in the peak/dip picture.   
Solid curves are Equations (\ref{zeta1}), (\ref{zeta2}), (\ref{zeta3}),  while histograms are obtained from 500,000 mock realizations via the algorithm described in Section \ref{pdesa}.
Each panel displays the numerical and theoretical expectations for the mean values and variances of these distributions; 
their agreement is excellent.}
\label{fig_p_eigen}
\end{figure}

The previous formulas  show explicitly that
the conditional distributions of shear eigenvalues are
Doroshkevich-like expressions, with shifted mean and reduced
variance.  Different panels in Figure  \ref{fig_p_eigen} display plots of these distributions, contrasted with 
results from the algorithm presented in Section \ref{pdesa}, where 500,000 mock realizations are considered.  
In particular, note the symmetry between $p(\zeta_1|\gamma)$ and
$p(\zeta_3|\gamma)$. 
We expect the mean values and variances of those distributions to be:
\begin{eqnarray}
\langle \zeta_1|\gamma \rangle &=& {3
  \over \sqrt{10 \pi}}\sigma_{\rm T} \sqrt{1-\gamma^2},~~~~ \sigma^2_{\rm \zeta_1|\gamma} = {13 \pi -27
  \over 30 \pi} \sigma^2_{\rm T}(1-\gamma^2)\\
  \langle \zeta_2|\gamma \rangle &=& 0,~~~~~~~~~~
\sigma^2_{\rm \zeta_2|\gamma} = {2 \over 15} \sigma^2_{\rm T} (1-\gamma^2)\\ 
\langle \zeta_3|\gamma \rangle &\equiv& - \langle \zeta_1|\gamma\rangle =
-{3 \over \sqrt{10 \pi}} \sigma_{\rm T} \sqrt{1-\gamma^2},~~ \sigma^2_{\rm \zeta_3|\gamma} \equiv  \sigma^2_{\rm \zeta_1|\gamma} 
= {13 \pi -27 \over 30 \pi} \sigma^2_{\rm T}(1-\gamma^2).
\end{eqnarray}
In the various panels, both theoretical and numerical expectations are reported and found to be in excellent agreement. 
For example, when $\gamma =0.50$, $\langle \zeta_1|\gamma \rangle_{\rm num}=0.4644$ while the expected theoretical value is 
 $\langle \zeta_1|\gamma \rangle_{\rm th}=0.4635$, and for the variances
 $\sigma^{2, \rm num}_{\zeta_1|\gamma} = 0.1102$
where the theoretical expectation is $\sigma^{2, \rm th}_{\zeta_1|\gamma} = 0.1101$. 

Obviously, to see explicitly the effect of the peak constraint,
 one needs to make cuts in $\xi_{\rm i}$ from the partial conditional distributions $\lambda_{\rm i}|\xi_{\rm i}$, since $\zeta_{\rm i} = \lambda_{\rm i} - \eta \xi_{\rm i} $.
In particular, expressions for  $p(\lambda_{\rm i}|{\bf H}>0, \gamma)$ can be derived from (\ref{zeta1}--\ref{zeta3}) via substitution of variables and
integration over ${\bf H}>0$. For example, it is straightforward to obtain an analytic formula for $p(\lambda_3|{\bf H}>0, \gamma)$.
In Section \ref{epd}, we will return on these issues in more detail, in connection with the conditional ellipticity and prolateness for dark matter halos.

\begin{figure}
\begin{center}
\includegraphics[angle=0,width=1.00\textwidth]{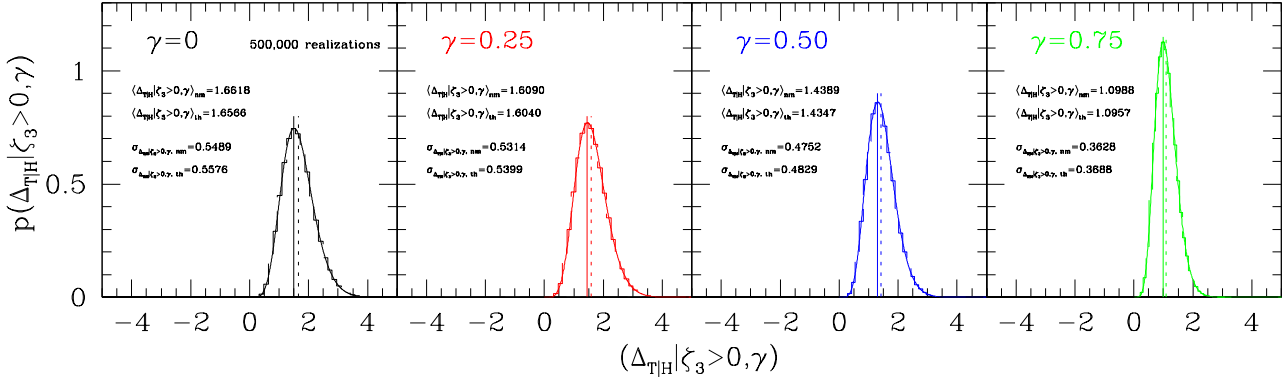}
\caption{Conditional probability distribution  $p(\Delta_{\rm T|H}|\zeta_3 >0, \gamma)$. Solid lines are obtained from
Equation (\ref{K1_positive_eq}), for different values of the correlation parameter $\gamma$, while  
 histograms are drawn from 500,000 realizations via the algorithm described in Section \ref{pdesa}. Note the excellent agreement 
 between numerical results and theoretical predictions for the corresponding mean and rms values, reported in the various panels.
In addition, solid vertical lines show the expected maximum value of each distribution, while dotted lines represent their corresponding mean value.}
\label{fig_p_K1_positive}
\end{center}
\end{figure}

Another important distribution is $p(\Delta_{\rm T|H}|\zeta_3>0, \gamma)$, with
$\Delta_{\rm T|H} \equiv K_1 = \zeta_1 + \zeta_2 + \zeta_3$ being the sum of the constrained eigenvalues, namely the probability distribution of $\Delta_{\rm T|H}$ confined in the regions with
$\zeta_3 > 0$ (i.e. all positive eigenvalues):
\begin{eqnarray}
\label{K1_positive_eq}
p(\Delta_{\rm T|H} |\zeta_3 >0, \gamma)
&=& -{75 \sqrt{5} \over 8 \pi \sigma^2_{\rm T} } {\Delta_{\rm T|H} \over (1-\gamma^2)}~{\rm exp} \Big( - {9 \over 8}
  {\Delta_{\rm T|H}^2 \over \sigma^2_{\rm T}(1-\gamma^2)} \Big) +  \\
&+& {25 \over 4 \sqrt{2 \pi} } {1 \over
  \sigma_{\rm T}\sqrt{1-\gamma^2}}~{\rm exp} \Big ( -
  {\Delta_{\rm T|H}^2 \over 2\sigma^2_{\rm T}(1-\gamma^2)} \Big ) \Big[{\rm erf}\Big ({\sqrt{10} \Delta_{\rm T|H}
  \over 4 \sigma_{\rm T}\sqrt{1-\gamma^2}} \Big )+ {\rm erf}\Big ({\sqrt{10} \Delta_{\rm T|H}
  \over 2 \sigma_{\rm T} \sqrt{1-\gamma^2}} \Big )    \Big]. \nonumber
\end{eqnarray}
This distribution is expected to have mean value
\begin{equation}
\langle \Delta_{\rm T|H} |\zeta_3 > 0, \gamma \rangle = {25 \sqrt{10} \over 144
  \sqrt{\pi}} (3 \sqrt{6} -2)~\sigma_{\rm T} \sqrt{1-\gamma^2}\simeq 1.66~\sigma_{\rm T}\sqrt{1-\gamma^2}.
\label{K1_positive_ave}
\end{equation}
It is also easy to see that the maximum of $p(\Delta_{\rm T|H}|\zeta_3>0, \gamma)$
is reached when $\Delta_{\rm T|H} \simeq 1.5~\sigma_{\rm T}\sqrt{1-\gamma^2}$, and 
with a more elaborate calculation its variance can be estimated analytically. The result is:
\begin{equation}
\sigma^2_{\Delta_{\rm T|H}|\zeta_3>0, \gamma}= {25 \over 4 \pi} \sigma^2_{\rm T} (1-\gamma^2)\Big[ {\rm arctan}(\sqrt{5}/2) +{\rm arctan}(\sqrt{5}) -{11 \sqrt{5} \over 54}   
-{125 \over 2592} (3 \sqrt{6}-2 )^2
\Big ] \simeq 0.31~\sigma^2_{\rm T} (1-\gamma^2).
\label{K1_positive_sig}
\end{equation}
Figure \ref{fig_p_K1_positive} confirms the previous relations, by  contrasting Equations (\ref{K1_positive_eq}), (\ref{K1_positive_ave})
and (\ref{K1_positive_sig}) with numerical results from 500,000 realizations from the algorithm presented in Section \ref{pdesa}.  
We find good agreement, and recover correctly the expected mean values and variances of the distributions. 
For example, when $\gamma=0.25$, we find   
$\langle \Delta_{\rm T|H}|\zeta_3>0, \gamma \rangle_{\rm num}=1.6090$ while the expected theoretical value is 
 $\langle \Delta_{\rm T|H}|\zeta_3>0, \gamma \rangle_{\rm th}=1.6040$, and for the rms values we
 find $\sigma^{\rm num}_{\Delta_{\rm T|H}|\zeta_3>0, \gamma} = 0.5314$
where the expected theoretical value is  $\sigma^{\rm th}_{\Delta_{\rm T|H}|\zeta_3>0, \gamma} = 0.5399$.
In the absence of correlations between the potential and
density fields (i.e. when $\gamma=0$), all the previous expressions reduce
consistently to the unconditional limit of Lee \& Shandarin (1998).

\subsection{Distribution of peak heights}

An important quantity which plays a major role in peaks theory (Bardeen et al. 1986)
is the distribution of peak heights, namely $p(\nu |{\bf H}>0, \gamma)$, where
$\nu = \delta_{\rm T}/\sigma_{\rm T}$ and the overdensity $\delta_{\rm T}$ is the trace of the shear tensor -- defined in Section \ref{jde}.
One can easily obtain a simplified analytic expression for this distribution as follows. 
First, normalize Equation (\ref{eigen_doro_ext}) by  $\sigma_{\rm T}$, defining  $N=\Delta_{\rm T|H} / \sigma_{\rm T}$
and $x= \delta_{\rm H}/ \sigma_{\rm H}$, so that $N = \nu - \gamma x$. In this framework, $x$ is the peak curvature of Bardeen et al. (1986) if Gaussian filters are used for 
computing the moments of the smoothed power spectrum (Equation \ref{eq_sig}).
Since $p(N|\gamma)$ is 
simply a Gaussian with mean zero and variance $(1-\gamma^2)$ -- see Rossi (2012) -- and 
because of Equation (\ref{eigen_doro_ext}), then
clearly $p(\nu|x, \gamma)$ is also Gaussian, with mean $ \gamma x$ and
reduced variance $(1-\gamma^2)$. Note in fact that $p(\nu,x|\gamma)$ is a bivariate Gaussian, because both $\nu$
and $x$ are  normally distributed random variates. This implies that $\langle \nu|x, \gamma\rangle = \gamma x$, and clearly $\langle \nu|{\bf H}>0, \gamma\rangle = 
\gamma \langle x| \xi_3>0 \rangle \equiv  \gamma \langle \nu | \lambda_3>0 \rangle$ (Bardeen et al. 1986) -- where $\langle \nu | \lambda_3>0 \rangle = 1.6566$ according to 
Equation (21) of Lee \& Shandarin (1998). 
Starting from a bivariate Gaussian with the previous expected mean value, 
it is then direct to derive a fairly
good approximation for the distribution of peak heights, namely
\begin{equation}
\label{eq_peak_heights_proxy}
p(\nu|{\bf H}>0, \gamma) = {1 \over  \sqrt{2 \pi}}~ {\rm exp} \Big [   { \-(\nu - \chi)^2 \over 2}  \Big ] 
\Big [ 1 + {\rm erf} \Big ( { \gamma (\nu-\chi) \over \sqrt{2 (1-\gamma^2)}}   \Big ) \Big ]      
\end{equation}
where
\begin{equation}
\chi = { \sqrt{2} \over \sqrt{\pi}}  \Big [   {25 \sqrt{5} \over 144} (3 \sqrt{6} -2) -1  \Big ] \gamma \simeq 0.86 \gamma.
\end{equation}

In essence, $p(\nu|{\bf H}>0, \gamma)$ is a Gaussian with shifted mean modulated by the role of the peak curvature, and it consistently reduces to
a zero-mean unit-variance Gaussian distribution when $\gamma=0$ -- as expected.
Additionally, the previous distribution can be shown to be equivalent to $p(\nu) [\int_0^{\infty} p(\xi_3|\gamma, \nu) {\rm d} \xi_3/p(\xi_3>0)]$, where the term in square brackets
quantifies the effect of the peak constrain -- and so the scale at which the difference between peaks and random positions would appear.

\begin{figure}
\centering
\includegraphics[angle=0,width=0.75\textwidth]{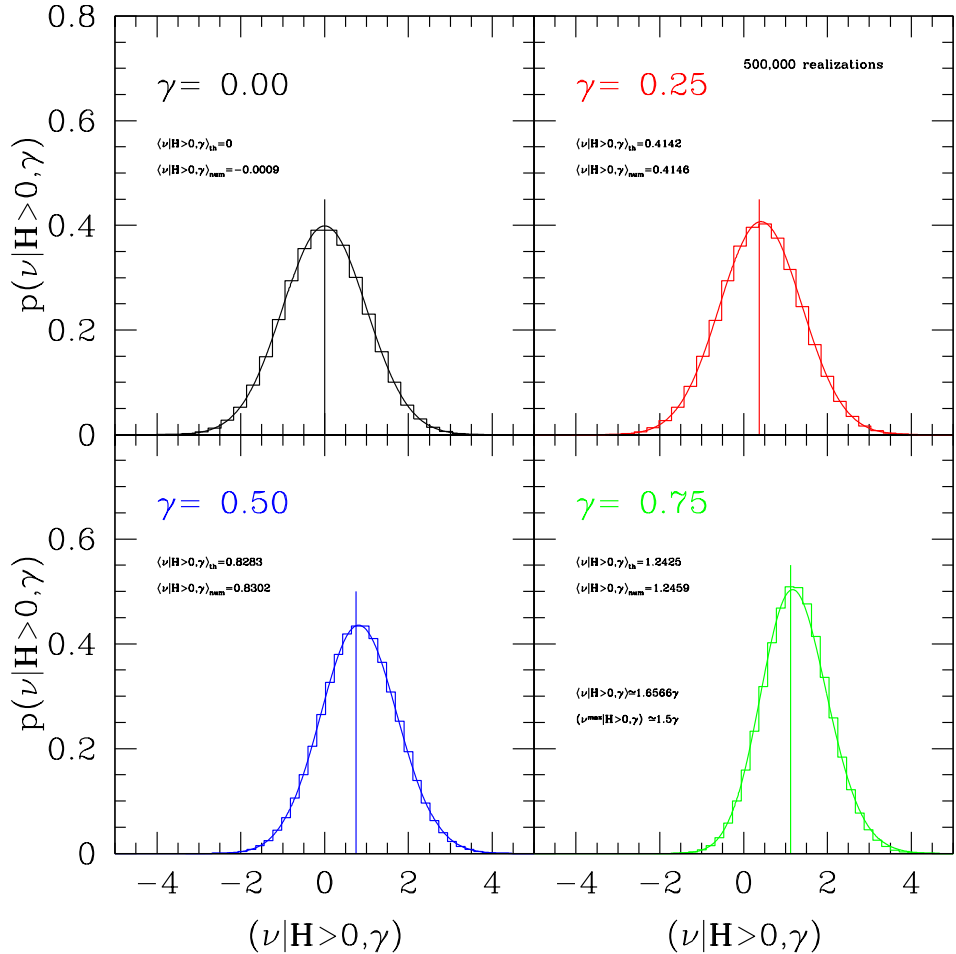}
\caption{Distribution of peak heights,  $p(\nu|{\bf H}>0, \gamma$), for different values of the correlation parameter $\gamma$ -- as specified in the various panels.
The numerical values for the mean of the distribution as a function of $\gamma$ are indicated in each panel, and also marked with vertical lines; they are in good agreement with theoretical expectations. 
Solid curves are drawn from Equation (\ref{eq_peak_heights_proxy}).}
\label{fig_peak_heights}
\end{figure}

Figure \ref{fig_peak_heights}
shows plots of $p(\nu|{\bf H}>0, \gamma)$ for different values 
of $\gamma$. In the various panels, vertical lines are the expected mean values;  we find good agreement 
between numerical estimates and theoretical expectations. For example, when $\gamma=0.25$, 
$\langle \nu|{\bf H}>0, \gamma \rangle_{\rm num} = 0.4146$ while $\langle \nu|{\bf H}>0, \gamma \rangle_{\rm th} = 0.4142$.
Solid curves in the figure are drawn from Equation (\ref{eq_peak_heights_proxy}), an approximation which is particularly good for lower values of the correlation strength.

\section{Conditional ellipticity and prolateness}
\label{epd}

The theoretical framework outlined in Section \ref{jde}, along with the algebra leading to the new computational procedure
presented in Section \ref{pdesa},
provides a natural way to include 
the peak/dip constraint into
the initial shape distributions of halos and voids.
This is a key aspect for
implementing an excursion set algorithm for peaks and dips in Gaussian random fields (which will be discussed in Section \ref{esmpd}), and has direct applications in 
the context of triaxial models of structure formation.
In particular, our description combines
the formalism of Bardeen et al. (1986) -- based on the density field  -- with that of Bond \& Myers (1996) -- focused on the shear field.

In what follows, first we introduce some useful definitions regarding shape parameters (ellipticity and prolateness); we then
provide a new expression for the 
joint conditional distribution of halo shapes given the density peak constraint, derived from Equation (\ref{doro_eigen}),  which generalizes some previous literature work.
The formula has important implications for the modeling of non-spherical dark matter halos, in relation to their initial shape distribution.  
Finally, we briefly discuss how the shear ellipticity and prolateness (and so the initial halo shapes) are modified by the inclusion of the density peak condition. 

Our analysis is based on the constrained eigenvalues of the initial shear field (Equations \ref{doro_eigen}--19), and we use the new algorithm to support our description.
While here we only consider the case for dark matter halos, it is straightforward to deal with voids. In a forthcoming companion publication,
we will present a more detailed study on the morphology of both halos and voids in the peak/dip picture, 
where we also investigate the modifications induced by primordial non-Gaussianity on their  shapes  (see also Section \ref{conc}).


\subsection{Shape parameters: general definitions}

In triaxial models of halo formation, such as the ellipsoidal collapse (Icke 1973; White \& Silk 1979; Barrow \& Silk 198; Kuhlman et al. 1996; Bond \& Myers 1996), 
it is customary to characterize the shape of a region by its ellipticity and prolateness. 
In particular, in the `peak-patch' approach of
Bond \& Myers (1996), the shape parameters are associated with the eigenvalues of the shear field ($\lambda_{\rm i}$ in our notation).
Considering the mapping ($\lambda_1, \lambda_2, \lambda_3 \rightarrow e_{\rm T}, p_{\rm T}, \delta_{\rm T}$), where the
subscript $T$ refers to the tidal tensor, we can write: 
\begin{equation}
\delta_{\rm T} \equiv k_1 = \lambda_1 + \lambda_2 + \lambda_3, \hspace{1cm} e_{\rm T} = {\lambda_1 -\lambda_3 \over 2 \delta_{\rm T}}, \hspace{1cm}
p_{\rm T} = {\lambda_1 + \lambda_3 - 2 \lambda_2  \over 2 \delta_{\rm T}} \equiv e_{\rm T} -{\lambda_2 -\lambda_3 \over \delta_{\rm T}},
\label{eq_dep_T}
\end{equation}
where $e_{\rm T}$ and $p_{\rm T}$ are the `unconditional' shear ellipticity and prolateness. 
In particular, if $\delta_{\rm T} > 0$ then $e_{\rm T} > 0$ and therefore $-e_{\rm T} \le p_{\rm T} \le  e_{\rm T}$. 

In the `peaks theory' of Bardeen et al. (1986), instead, ellipticity ($e_{\rm H}$) and prolateness ($p_{\rm H}$) 
are associated with the eigenvalues of the density field ($\xi_{\rm i}$ in our notation).
Hence,  given the mapping ($\xi_1, \xi_2, \xi_3 \rightarrow e_{\rm H}, p_{\rm H}, \delta_{\rm H}$)  where now the
subscript $H$ refers to the Hessian of the density field, one has: 
\begin{equation}
\delta_{\rm H} \equiv h_1 = \xi_1 + \xi_2 + \xi_3, \hspace{1cm} e_{\rm H} = {\xi_1 -\xi_3 \over 2 \delta_{\rm H}}, \hspace{1cm}
p_{\rm H} = {\xi_1 + \xi_3 - 2 \xi_2  \over 2 \delta_{\rm H}} \equiv e_{\rm H} -{\xi_2 -\xi_3 \over \delta_{\rm H}}.
\label{eq_dep_H}
\end{equation} 
As in the previous case, if $\delta_{\rm H} > 0$ then $e_{\rm H} > 0$ and $-e_{\rm H} \le p_{\rm H} \le  e_{\rm H}$. 

It is therefore natural to consider the `conditional' mapping ($\zeta_1, \zeta_2, \zeta_3 \rightarrow E_{\rm T|H}, P_{\rm T|H}, \Delta_{\rm T|H}$)
and define:
\begin{equation}
\Delta_{\rm T|H} \equiv K_1 = \zeta_1 +\zeta_2+\zeta_3, \hspace{1cm}
E_{\rm T|H} = {\zeta_1 - \zeta_3 \over 2 \Delta_{\rm T|H}}, \hspace{1cm} 
P_{\rm T|H} = {\zeta_1 + \zeta_3 - 2 \zeta_2 \over 2 \Delta_{\rm T|H} } \equiv E_{\rm T|H} - {\zeta_2 -\zeta_3 \over \Delta_{\rm T|H}},
\label{eq_dep_T_given_H}
\end{equation}
where $\zeta_{\rm i}$ are the constrained shear eigenvalues -- given by Equation (19).
In what follows, we will refer to $E_{\rm T|H}$ and  $P_{\rm T|H}$ as to the conditional shear ellipticity and prolateness, respectively; we will also
relate these quantities to the unconditional expressions previously introduced, involving ($e_{\rm T},p_{\rm T}$) and ($e_{\rm H},p_{\rm H}$).


\subsection{Joint conditional distribution of shear ellipticity and prolateness in the peak/dip picture}

Combining  (\ref{eq_dep_T}) with Doroshkevich's formula (\ref{doro_original}), it is straightforward
 to derive the `unconditional' distribution of $e_{\rm T}$ and $p_{\rm T}$ given $\delta_{\rm T}$, $g(e_{\rm T},p_{\rm T}|\delta_{\rm T})$, related to the 
gravitational potential. In particular,
\begin{equation}
g(\delta_{\rm T}, e_{\rm T}, p_{\rm T}) = p(\delta_{\rm T}) g( e_{\rm T}, p_{\rm T}| \delta_{\rm T}) 
\label{gep_uncond}
\end{equation}
where $p(\delta_{\rm T})$ is simply a Gaussian with zero mean and variance $\sigma_{\rm T}^2$, and
\begin{equation}
g(e_{\rm T}, p_{\rm T}| \delta_{\rm T}) = {15^3 \over 3 \sqrt{10 \pi} } \Big ( {\delta_{\rm T} \over \sigma_{\rm T} } \Big )^5 e_{\rm T} (e^2_{\rm T} - p^2_{\rm T}) 
~{\rm exp} \Big[ -{5 \over 2}  \Big({\delta_{\rm T} \over \sigma_{\rm T}} \Big)^2 (3 e^2_{\rm T} + p^2_{\rm T} )\Big]. 
\label{gep_uncond_bis}
\end{equation}
Equation (\ref{gep_uncond}) implies that 
the unconditional joint distribution of shear ellipticity and prolateness is independent of that of the overdensity $\delta_{\rm T}$.
Similarly, inserting (\ref{eq_dep_H}) in Doroshkevich's formula (\ref{doro_original_bis}), one obtains
\begin{equation}
g(\delta_{\rm H}, e_{\rm H}, p_{\rm H}) = p(\delta_{\rm H}) g( e_{\rm H}, p_{\rm H}| \delta_{\rm H}) 
\end{equation}
for the quantities related to the density Hessian,
where $p(\delta_{\rm H})$ is a Gaussian with zero mean and variance $\sigma_{\rm H}^2$, while
\begin{equation}
g(e_{\rm H}, p_{\rm H}| \delta_{\rm H}) = {15^3 \over 3 \sqrt{10 \pi} } \Big ( {\delta_{\rm H} \over \sigma_{\rm H} } \Big )^5 e_{\rm H} (e^2_{\rm H} - p^2_{\rm H}) 
~{\rm exp} \Big[ -{5 \over 2}  \Big({\delta_{\rm H} \over \sigma_{\rm H}} \Big)^2 (3 e^2_{\rm H} + p^2_{\rm H} )\Big]. 
\end{equation}

Along the same lines, combining (\ref{eq_dep_T_given_H}) with (\ref{doro_eigen}) and using the definitions introduced in the previous section, it is direct to obtain
\begin{equation}
G( \Delta_{\rm T|H}, E_{\rm T|H}, P_{\rm T|H}|\gamma) = p(\Delta_{\rm T|H}|\gamma) G(E_{\rm T|H},P_{\rm T|H}|\Delta_{\rm T|H}, \gamma)
\label{eq_epc_base}
\end{equation}
where $p(\Delta_{\rm T|H}|\gamma)$ is a Gaussian distribution with mean zero and variance $\sigma^2_{\rm T} (1 -\gamma^2)$ -- i.e. Equation (57) in Rossi (2012) --
and
\begin{equation}
G(E_{\rm T|H}, P_{\rm T|H}| \Delta_{\rm T|H}, \gamma) = {15^3 \over 3 \sqrt{10 \pi}} \Big ( {\Delta_{\rm T|H} \over \sigma_{\rm T}} \Big )^5 {1 \over (1 - \gamma^2)^{5/2}} E_{\rm T|H} (E_{\rm T|H}^2 - P_{\rm T|H}^2) 
~{\rm exp} \Big [ {5 \over 2 (1 -\gamma^2)} \Big ( {\Delta_{\rm T|H} \over \sigma_{\rm T}}\Big )^2 (3 E_{\rm T|H}^2 + P_{\rm T|H}^2)    \Big ].
\label{eq_epc}
\end{equation}
The previous expression (\ref{eq_epc}) is the joint conditional distribution of shear ellipticity and prolateness, given $\Delta_{\rm T|H}$ and a correlation strength $\gamma$ with the density field.
Clearly, $G(E_{\rm T|H}, P_{\rm T|H}| \Delta_{\rm T|H}, \gamma)$ is also independent of the distribution of  $(\Delta_{\rm T|H}|\gamma)$.
Equation (\ref{eq_epc}) generalizes the  standard joint distribution of halo shape parameters to include the peak constraint, and is another main result of this paper.
Note that, in the absence of correlation (i.e. when $\gamma=0$), $E_{\rm T|H} \rightarrow e_{\rm T}$,  $P_{\rm T|H} \rightarrow p_{\rm T}$,  $\Delta_{\rm T|H} \rightarrow \delta_{\rm T}$, so that
(\ref{eq_epc}) reduces consistently to the `unconditional' Doroshkevich's limit (\ref{gep_uncond_bis}).

From this joint conditional probability function, it is possible to derive the partial conditional distributions for the shear ellipticity and prolateness given the peak constrain (which we will discuss next),  
and their expressions at density peak locations, when the condition ${\bf H}>0$ is satisfied.


\begin{figure}
\centering
\includegraphics[angle=0,width=1.00\textwidth]{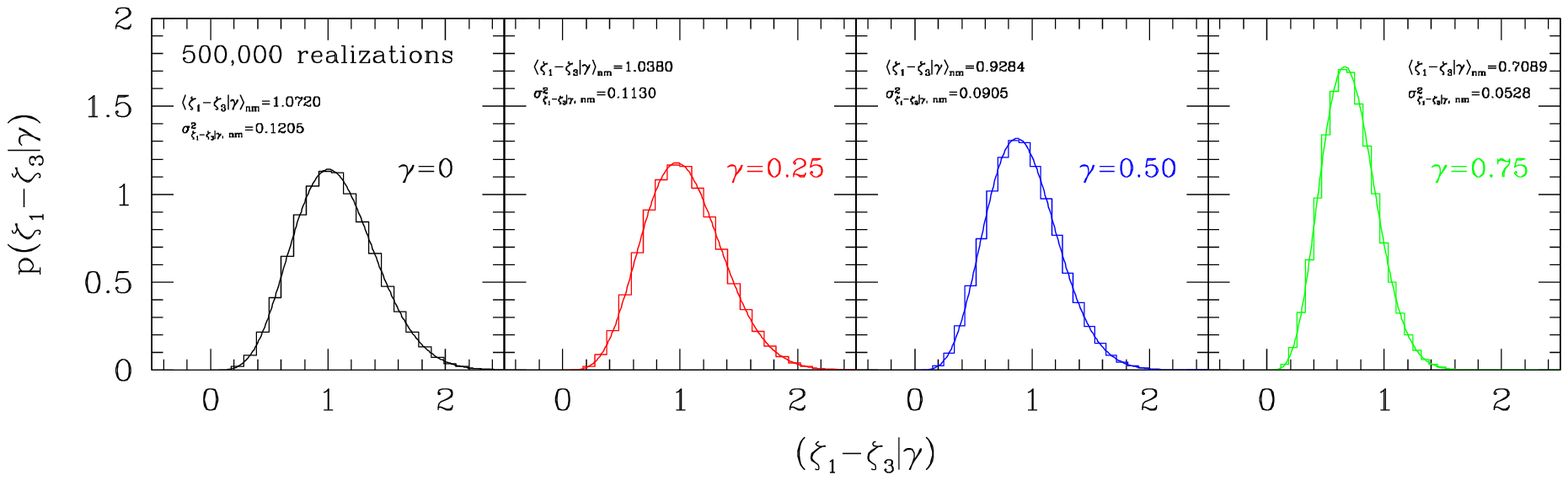}
\includegraphics[angle=0,width=1.00\textwidth]{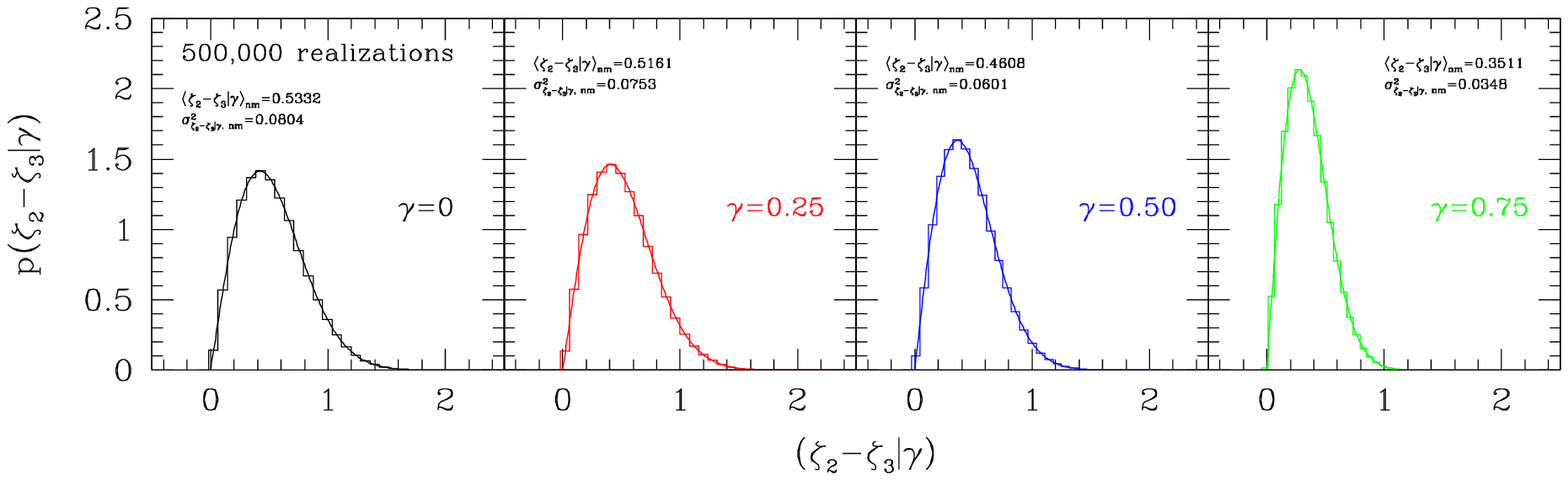}
\caption{[Top] Distribution of $(\zeta_1-\zeta_3|\gamma)$, for different values of the correlation strength $\gamma$ -- as indicated in the panels.
Its shape is directly related to the shear conditional ellipticity in the peak/dip picture -- see Equation (\ref{eq_ed_cond}).  
[Bottom] Distribution of $(\zeta_2-\zeta_3|\gamma)$, for different values of the correlation strength $\gamma$.
This distribution is instead related to the shear conditional prolateness in the peak/dip picture -- see Equation (\ref{eq_pd_cond}).  
The numerical values for the mean and variance of the distributions, as a function of $\gamma$, are also indicated in the figures.
In both cases, solid curves are obtained via numerical integrations starting from the joint conditional distribution of eigenvalues (\ref{doro_eigen}).
Histograms are drawn from 500,000 realizations, using the algorithm described in Section \ref{pdesa}.}
\label{fig_cep}
\end{figure}

\begin{figure}
\centering
\includegraphics[angle=0,width=1.00\textwidth]{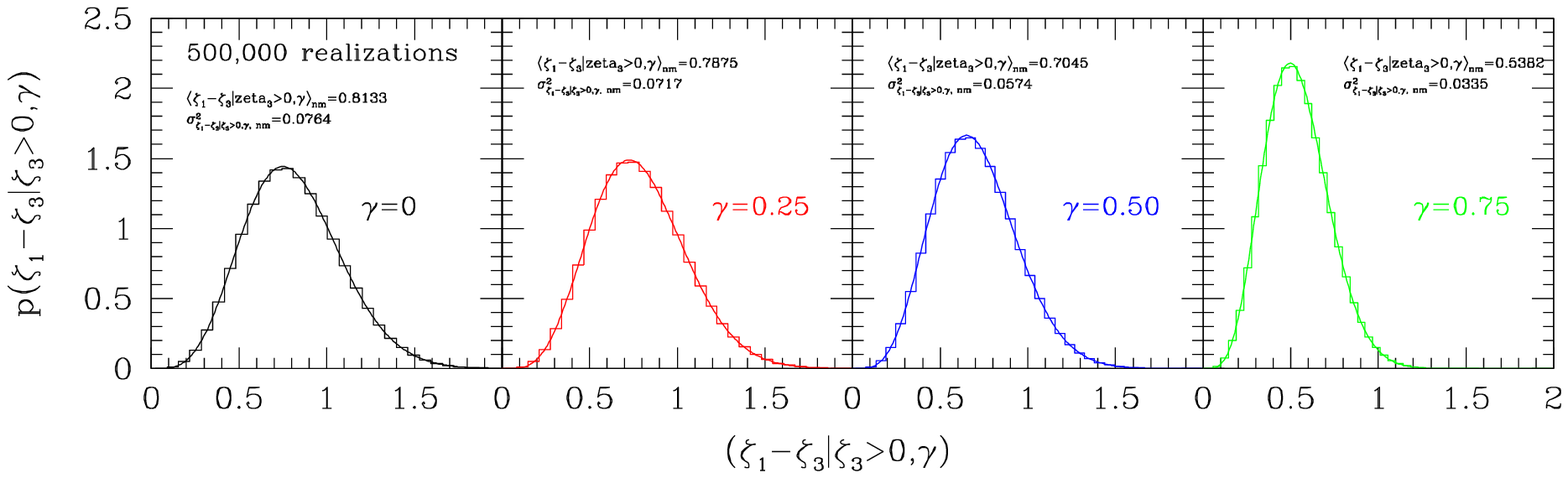}
\includegraphics[angle=0,width=1.00\textwidth]{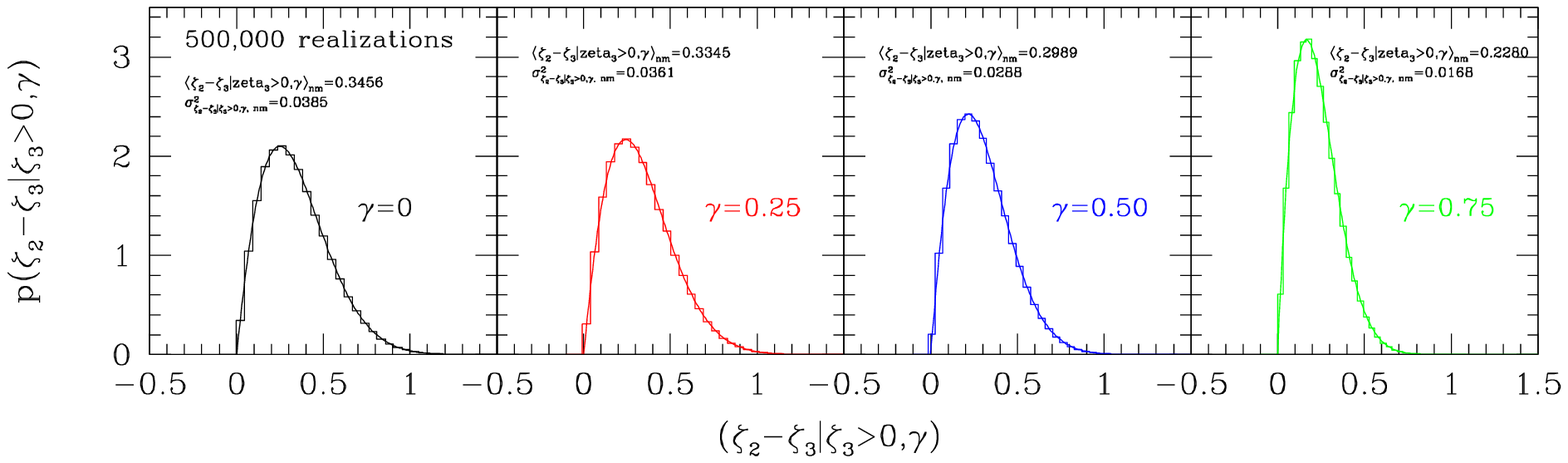}
\caption{Distributions $p(\zeta_1-\zeta_3|\zeta_3>0,\gamma)$ [top] and $p(\zeta_2-\zeta_3|\zeta_3>0,\gamma)$  [bottom]
for different values of the correlation strength $\gamma$.
In both figures, histograms are drawn from 500,000 realizations using the algorithm described in Section \ref{pdesa}, while 
solid curves are obtained via numerical integrations of the joint conditional distribution of eigenvalues (\ref{doro_eigen}).
In the various panels, we also provide the numerical values for the mean and variance of the distributions, as a function of $\gamma$. 
In particular, their mean values at $\gamma=0$ allow one to
quantify the shift in the mean of the shear ellipticity and prolateness caused by the peak constraint.}
\label{fig_cep_pos}
\end{figure}


\subsection{Initial distributions of  triaxial halo shapes at density peak locations}

We can readily express the conditional distributions of shear ellipticity and prolateness in the peak/dip picture in terms of
the unconditional quantities (\ref{eq_dep_T}) and (\ref{eq_dep_H}). This is done simply by recalling that  the 
constrained shear eigenvalues are given by $\zeta_{\rm i} = \lambda_{\rm i} - \eta \xi_{\rm i}$, according to Equation (19).

 It is then direct to obtain:
\begin{equation}
E_{\rm T|H} \Delta_{\rm T|H}  = {(\zeta_1 - \zeta_3) \over 2} \equiv {(\lambda_1 -\lambda_3) \over 2} - \eta {(\xi_1 - \xi_3) \over 2} =  \delta_{\rm T} e_{\rm T} - \eta \delta_{\rm H} e_{\rm H}.
\label{eq_ed_cond}
\end{equation}
The previous relation implies that, for a given $\Delta_{\rm T|H}$, the conditional shear ellipticity will have its mean value shifted by 
the presence of the peak constraint; the entity of the shift has to be ascribed to the additional factor $\eta \delta_{\rm H} e_{\rm H}$, which is given by the density ellipticity 
(essentially, the latter term quantifies the role of the peak curvature).  
The top panel of Figure \ref{fig_cep} shows the distribution of $(\zeta_1-\zeta_3|\gamma)$ for different values of the correlation strength $\gamma$, 
namely the combination of constrained shear eigenvalues which controls the conditional ellipticity 
-- according to Equation (\ref{eq_ed_cond}). 
Histograms are drawn from 500,000 realizations using the algorithm described in Section \ref{pdesa}, while 
solid curves are obtained via numerical integration -- starting  from the joint conditional distribution of eigenvalues (\ref{doro_eigen}).
In the various panels, we also provide the numerical values for the mean and variance of the distribution, as a function of $\gamma$. 

Similarly, 
\begin{equation}
P_{\rm T|H}  \Delta_{\rm T|H}  = {(\zeta_1 + \zeta_3 -2 \zeta_2)\over 2} =  {(\lambda_1 + \lambda_3 -2 \lambda_2)\over 2} -\eta {(\xi_1 + \xi_3 -2 \xi_2)\over 2}  = \delta_{\rm T} p_{\rm T} - \eta \delta_{\rm H} p_{\rm H}
 \equiv E_{\rm T|H}  \Delta_{\rm T|H}  - (\zeta_2 -\zeta_3). 
\label{eq_pd_cond}
\end{equation}
This expression shows that the conditional shear prolateness can be obtained from the conditional shear ellipticity (\ref{eq_ed_cond}) and the
combination ($\zeta_2 - \zeta_3$) of the constrained shear eigenvalues. 
The bottom panel of Figure \ref{fig_cep} displays the distribution of $(\zeta_2-\zeta_3|\gamma)$,
for different values of the correlation strength $\gamma$.  Again, solid curves are derived from a numerical integration of  Equation (\ref{doro_eigen}).
 
Of particular interest are also the distributions $g(e_{\rm T}|{\bf H}>0, \gamma)$ and $g(p_{\rm T}|{\bf H}>0, \gamma)$, which will provide
the initial triaxial shapes of dark matter halos at peak locations. They
are related to the previous expressions, and can be readily obtained  
within the outlined formalism; clearly, they will reduce to the standard (or unconditional) shear ellipticity and
prolateness when $\gamma=0$, in the absence of correlation.
We will present a dedicated study focused on the morphology of halos and voids in the peak/dip picture, where we characterize in detail those
distributions, especially in relation to the ellipsoidal collapse model.  We anticipate here that several analytic and insightful results on their shapes
can be derived, starting from (\ref{eq_epc}) and using (\ref{doro_eigen}), (\ref{eq_ed_cond}) and (\ref{eq_pd_cond}).   
For example, according 
to (\ref{eq_ed_cond}), when ${\bf H} >0$ (which is equivalent to impose the condition $\xi_3>0$ on an ordered set of density Hessian eigenvalues), 
then the quantity $\langle \xi_1 -\xi_3| \xi_3 >0  \rangle$
will essentially provide by how much the mean value of the shear ellipticity has been shifted by the peak constrain. 
For the prolateness the situation is slightly more complicated, but according to (\ref{eq_pd_cond}) 
the shift can be derived by the additional knowledge of  $\langle \xi_2 -\xi_3| \xi_3 >0  \rangle $.
To this end,  Figure \ref{fig_cep_pos} shows the distributions of $(\zeta_1-\zeta_3|\zeta_3>0,\gamma)$ and $(\zeta_2-\zeta_3|\zeta_3>0,\gamma)$, 
for different values of the correlation strength $\gamma$ (note that the quantities just discussed are their mean values when $\gamma=0$).  
As in the previous plot, histograms are drawn from 500,000 realizations using the algorithm described in Section \ref{pdesa}, while 
solid curves are obtained via numerical integrations -- starting  from the joint conditional distribution of eigenvalues (\ref{doro_eigen}).
In the various panels, we also provide the numerical values for the mean and variance of the distribution, as a function of $\gamma$. 

As mentioned before, we will return in more detail on these distributions in a companion publication, to quantify the impact of the peak
constrain on their shapes. We will also make the connection
with the work of Bardeen et al. (1986) more explicit -- see their Appendix C. 
It will be also interesting to include the role of the peculiar gravity field in our description, along the
lines of van de Weygaert \& Bertschinger (1996), as well as to extend the work of Desjacques (2008) on the joint statistics of the shear tensor
and on the dynamical aspect of the environmental dependence, within this formalism. 



\section{Excursion set approach for peaks and dips in Gaussian random fields}
\label{esmpd}


The ability to distinguish between random positions and peaks/dips, contained in Equations (\ref{doro_inter_extended}) or (\ref{doro_eigen}) 
and achieved by the algorithm presented in Section \ref{pdesa}, is the key to implement
an excursion set model for peaks and dips in Gaussian random fields.
Indeed, the primary motivation (and one of the main strengths) of the proposed sampling procedure resides in its direct inclusion
into the excursion set framework (Epstein 1983; Peacock \& Heavens 1990; Bond et al. 1991; Lacey \& Cole 1993).
In essence, because in our formalism the peak overdensity is simply the trace of the conditional shear tensor (recall the definitions in Section \ref{jde}),  
and since in triaxial models of collapse the initial shape parameters are just combinations of the shear eigenvalues  (Bond \& Myers 1996), 
our prescription provides a direct way to generate the distribution of initial overdensities under the conditions that they are peaks/dips (i.e. the distribution of
peak heights, when ${\bf H}>0$) -- along with the corresponding conditional distribution of initial shapes (see Section \ref{epd}). 
In this respect, it is then straightforward to include this part in standard excursion set algorithms, as those used for example in Chiueh \& Lee (2001),  Sheth \& Tormen (2002) 
or Sandvik et al. (2007)
to compute the mass function, or
in Rossi, Sheth \& Tormen (2011) to describe halo shapes; the `peak/dip excursion-set-based' algorithm is also useful for computing the halo bias assuming triaxial models of structure formation (i.e. ellipsoidal collapse). 
The only main conceptual difference is the pre-selection of peak/dip locations, instead of random positions in the field. The mass scale of the peak will then be fixed
by finding the proper $\sigma_{\rm T}$ which satisfies the combination of $(\delta_{\rm T}, e_{\rm T}, p_{\rm T}|{\bf H}>0, \gamma)$, assuming some structure formation models -- as for example the ellipsoidal collapse.

This is particularly useful because the excursion set theory is a powerful
tool for understanding  various aspects of the full dynamical complexity of halo formation.
Perturbations are assumed to evolve stochastically with the smoothing scale, and the 
problem of computing the probability of halo formation is mapped into the
classical first-passage time problem in the presence of a barrier.
A very elegant reformulation of this theory has been recently proposed by 
Maggiore \& Riotto (2010a,b,c), who made several technical and 
conceptual improvements (i.e. no ad hoc absorbing barrier boundary conditions, account for non-markovianity induced by filtering,  
unambiguous mass association to a smoothed scale, etc.)
by deriving the original excursion set theory from a path integral formulation -- following a microscopical approach.
These authors also noted that the failure of the standard excursion set approach 
may be related to the inadequacy of the oversimplified physical model adopted 
 for halo formation (either spherical or ellipsoidal), and propose to 
 treat the critical threshold for collapse as a stochastic variable
which better captures some of the dynamical complexity of the halo formation phenomenon.
Even so, they find that the non-markovian contributions do not alleviate the discrepancy between excursion set predictions and $N$-body simulations.

In addition to the problems pointed out by Maggiore \& Riotto (2010a,b,c), there is also the fact that 
-- in its standard formulation -- the  excursion set approach is unable to differentiate between peaks/dips and random locations in space -- i.e. all points are
treated equally. However, since  local extrema are plausible sites for the formation of
nonlinear structures and there is a 
good correspondence 
between peaks in the initial conditions and halos at late times,
it may be important to differentiate between those special positions in space. The algorithm proposed here goes in this
direction, since it allows one to pre-select those special points in space around which halos form (peaks), and not just random locations -- and permits to associate their corresponding
 initial  shape distribution:  in essence, the computational procedure selects a special subset, among all the possible random walks considered in the standard excursion set procedure. 

Therefore, the `peak/dip excursion-based' algorithm
can be used to study the mass function of halos and their triaxial  shapes at peak/dip positions, and also the halo bias. 
In fact, our prescription allows one to generate the initial distributions of overdensity, ellipticity and prolateness
(i.e. shape parameters) at a scale set by the variance $\sigma_{\rm T}$, with the constraint that $\delta_{\rm T}$ is a peak (i.e. the condition ${\bf H}>0$ on the Hessian).
One can then just evolve this initial conditional shape distribution by solving a dynamical equation of collapse, and 
study the final shape distribution (as done for example in Rossi, Sheth \& Tormen 2011) but now at peak/dip locations.
Alternatively, one can adopt a `crossing threshold', since in the excursion set approach
an halo is formed when the smoothed density perturbation reaches a critical value for the first time, and 
the problem is reduced to a first-passage problem  in the presence of a barrier (i.e. if the overdensity exceeds a critical value, the random
walk stops at this scale; if not, the walk continues to smaller scales).
Moreover,  our numerical technique can be easily integrated in 
Montecarlo realizations of the trajectories obtained from a Langevin equation with colored noise (i.e. Bond et al. 1991; Robertson et al. 2009)
at peak/dip locations, and even for situations where the walks are correlated -- in presence of non-markovian effects, along the lines of De Simone et al. (2011a,b).
In this respect, our technique is general, since any kind of filter function can be readily implemented. 
All these lines of research are ongoing efforts,  and will be presented in several forthcoming publications. 



\section{Conclusions}
\label{conc}

From the joint conditional probability distribution of an ordered set of shear eigenvalues
in the peak/dip picture 
(Rossi 2012; Section \ref{jde}, Equations \ref{doro_inter_extended} and \ref{doro_eigen}), we derived a new
algorithm to sample the constrained eigenvalues of the initial shear field associated with Gaussian statistics at peak/dips
positions in the correlated density field. 
The algorithm, described in Section \ref{pdesa}, was then used 
to test and confirm several formulas presented in Rossi (2012) regarding conditional distributions and probabilities, subjected to the extremum constraint
(Section \ref{cdpnt}). We found excellent agreement between numerical results and theoretical predictions (Figures \ref{fig_p_eigen} and \ref{fig_p_K1_positive}).
In addition, we also showed how the standard distributions of shear ellipticity and prolateness 
in triaxial models of structure formation
are modified by the constraint (Section \ref{epd}; Figures \ref{fig_cep} and \ref{fig_cep_pos}), and 
provided a new expression for the 
conditional distribution of shape parameters given the density peak requirement (Equation \ref{eq_epc}), which generalizes some previous literature work.
The formula has important implications for the modeling of non-spherical dark matter halo shapes, in relation to their initial shape distribution, and is directly applicable to the
ellipsoidal collapse model (Icke 1973; White \& Silk 1979; Barrow \& Silk 1981; Kuhlman et al. 1996; Bond \& Myers 1996). In particular, our novel description is able to combine consistently, for the first time,
the formalism of Bardeen et al. (1986) -- based on the density field -- with that of Bond \& Myers (1996) -- based on the shear field.
Along the way,  we also discussed the distribution of peak heights (see Figure \ref{fig_peak_heights}), 
which plays a major role in peaks theory (Bardeen et al. 1986).

While the primary motivation of this paper was to illustrate the new `peak/dip excursion-set-based' algorithm, and to show a few applications focused on the morphology of the cosmic web 
(following up, and complementing with some more insights, the theoretical work 
presented in Rossi 2012), the other goal was to  
describe how the new sampling procedure naturally integrates into the standard excursion set framework (Epstein 1983; Peacock \& Heavens 1990; Bond et al. 1991; Lacey \& Cole 1993) -- 
potentially solving some of its well-known problems.
In particular, in Section \ref{esmpd} we argued 
that the ability to distinguish between random positions and peaks/dips, encoded in the algorithm and in Equations (\ref{doro_inter_extended}) and (\ref{doro_eigen}) derived from first principles,   
is indeed the key to implement a generalized excursion set model for peaks and dips in Gaussian random fields.
This is the actual strength of the proposed computational procedure, since part of the failure of the original excursion set theory
may be attributed to its inability to differentiate between random 
positions and special points (peaks) in space around which halos may form.

To this end, our simple prescription can be used to study the halo mass function, halo/void shapes and bias
at peak/dip density locations, in conjunction with triaxial models of structure formation. All these research lines are ongoing efforts,  subjects of several forthcoming publications. 
The essential part is the characterization of the distribution of peak heights, and of the initial shape distribution at peak/dip locations (Equations \ref{doro_inter_extended}, \ref{doro_eigen}, and \ref{eq_epc}).

The algorithm presented in this paper offers also a much broader spectrum of applications.
This is because, as pointed out by Rossi (2012), the fact that the eigenvalues of the Hessian matrix can be used to
discriminate between different types of structures in a particle distribution
is fundamental to a number of structure-finding algorithms, shape-finders algorithms, 
and structure reconstruction on the basis of tessellations. 
For example, it can be used for studying the dynamics and morphology of cosmic voids -- 
see for example van de Weygaert \& Platen (2011), Bos et al. (2012), Pan et al. (2012),
and
the Monge-Amp\`{e}re-Kantorovitch reconstruction procedure
by Lavaux \& Wandelt (2010) -- and in
several observationally-oriented applications, in
particular for developing 
algorithms to find and classify structures in the cosmic web or in relation to its
skeleton (Sahni et
al. 1998; Schaap \&
van de Weygaert 2000; Novikov et al. 2006; Hahn et al. 2007;
Romano-Diaz \& van de Weygaert 2007; Forero-Romero et al. 2009;  Zhang et al. 2009; Lee \& Springel 2010; 
Arag{\'o}n-Calvo et al. 2010a,b; Platen et al. 2011; Cautun et al. 2012; Hidding et al. 2012). 
Another application is related to the work of
Bond, Strauss \& Cen (2010), who presented an algorithm that
uses the eigenvectors of the Hessian matrix of the smoothed galaxy
distribution to identify individual filamentary structures.
In addition, since galaxy clusters are related to primordial density peaks, 
and there is  a correspondence between structures in the evolved density field and local properties
of the linear tidal shear, 
our theoretical framework provides a direct way to relate initial conditions and observables from galaxy clusters. 

Other intriguing connections involve topological studies of the cosmic web, the genus statistics and
Minkowski functionals (Gott et al. 1986, 1989; Park et al. 1991, 2005; Matsubara 2010), and the possibility to address open questions regarding the origin of angular momentum and halo spin within this framework;
this is because the dependence of the spin alignment on the morphology of the large-scale 
mass distribution is due to the difference in shape of the tidal 
fields in different environments, and most of the halo properties depend significantly on environment, and in particular on the tidal field
-- i.e.  the environmental dependence of halo assembly time and unbound substructure fraction has
its origin from the tidal field
(Wang et al. 2011). 
It will be also interesting to explore how the new formalism proposed here
can be used to study halo spin, shape and the orbital angular momentum of subhaloes relative to the LSS, in the context of 
the eigenvectors of the velocity shear tensor (see the recent study by Libeskind et al. 2013). 
In addition, the more complex question of the
local expected density field alignment/orientation distribution as a
function of the local field value (Bond
1987; Lee \& Pen 2002; Porciani et al. 2002;  Lee, Hahn \& Porciani 2009; Lee 2011) can be addressed within this framework, and is the subject 
of future studies. 

On the theoretical side, we note that our algorithm is 
 restricted to one scale (i.e. peaks and dips in the density field, as in Bardeen et al. 1986), but
the extension to a multiscale \textit{peak-patch} approach along the lines of Bond \& Myers
(1996) is doable and subject of ongoing work. As argued in Rossi (2012), this will 
allow to account for the role of the peculiar
 gravity field itself, an important aspect not considered in our formalism but discussed in detail in 
 van de Weygaert \& Bertschinger (1996).
 Including all these effects in our framework and translating them into a more elaborated algorithm
 is ongoing effort, and will allow us to make the connection with 
the  multiscale analysis of the Hessian
matrix of the density field by 
van de Weygaert \& Bertschinger (1996) and
Arag{\'o}n-Calvo et al. (2007;  2010a,b).
It will also allow us to incorporate the distortion effect of the peculiar gravity field in our initial distribution of halo/void shapes (Section \ref{epd}). 
The natural extension of the peak/dip picture for the initial shear to non-Gaussian fields is also ongoing effort, along with some other broader applications in  the context of the excursion set model -- 
for example in relation to the hot and cold spots in the Cosmic Microwave Background, including the effects of $f_{\rm NL}$-type non-Gaussianity on their shapes (i.e. Rossi, Chingangbam \& Park 2011) -- which will be presented in forthcoming publications.



\section*{Acknowledgments}

The final stage of this work was completed 
during the `APCTP-IEU Focus Program on Cosmology and Fundamental Physics III'  (June 11-22, 2012) at Postech, in Pohang, Korea. 
I would like to thank the organizers of the workshop, and in particular Changrim Ahn.
Also, many thanks to Changbom Park for a careful reading of the
manuscript, and for many interesting 
discussions, suggestions and encouragement.





\appendix


\section{Essential notation} \label{appendix_notation}
\label{notation}


We summarize here the basic notation adopted in the paper, which is  essentially the same as the one introduced by Rossi (2012) -- with a few minor changes to make the
connection with previous literature more explicit.
In particular, let $\Psi$ denote the displacement field, $\Phi$ the potential of the displacement field (i.e. the gravitational field), 
$F$ the source of the displacement field (i.e. the density field).
Both $F$ and $\Phi$ are Gaussian random fields, the latter specified by the matter power spectrum $P(k)$, with $k$
denoting the wave number and $W(k)$ the smoothing kernel.
Use $T_{\rm ij}$ for the shear tensor (its eigenvalues are $\lambda_1, \lambda_2, \lambda_3$),
$H_{\rm ij}$ for the Hessian matrix of the density $F$ (with eigenvalues  $\xi_1, \xi_2, \xi_3$), $J_{\rm ij}$ for the Jacobian of the
displacement field, where $i,j=1,2,3$.  
Indicate with ${\bf q}$ the Lagrangian coordinate, with ${\bf x}$ the Eulerian coordinate, where
${\bf x} ({\bf q}) = {\bf q} + \Psi({\bf q})$.
Clearly, 
\begin{equation}
J_{\rm ij}({\bf q}) = { \partial x_{\rm i}  \over \partial q_{\rm j}}
= \delta_{\rm ij} + T_{\rm ij}, \hspace{1cm}  
T_{\rm ij} =
{\partial^2 \Phi  \over \partial q_{\rm i} \partial q_{\rm j}},  \hspace{1cm} 
H_{\rm ij} = {\partial^2  F  \over \partial q_{\rm i} \partial
  q_{\rm j}}, \hspace{1cm} 
  F ({\bf q}) = \sum_{\rm i=1}^3 {\partial \Psi_{\rm i} \over
  \partial q_{\rm i}} \equiv  \sum_{\rm i=1}^3 {\partial^2 \Phi \over
  \partial q^2_{\rm i}}.
\end{equation}
The correlations between the shear and density Hessian are expressed by:
\begin{equation}
\langle   T_{\rm ij} T_{\rm kl} \rangle = { \sigma_{\rm T}^2 \over 15}
( \delta_{\rm ij} \delta_{\rm kl}  + \delta_{\rm ik} \delta_{\rm jl} +
\delta_{\rm il} \delta_{\rm jk})  
\end{equation}

\begin{equation}
\langle   H_{\rm ij} H_{\rm kl} \rangle = { \sigma_{\rm H}^2 \over 15}
( \delta_{\rm ij} \delta_{\rm kl}  + \delta_{\rm ik} \delta_{\rm jl} +
\delta_{\rm il} \delta_{\rm jk})  
\end{equation}
 
\begin{equation}
\langle   T_{\rm ij} H_{\rm kl} \rangle = { \Gamma_{\rm TH} \over
  15} ( \delta_{\rm ij} \delta_{\rm kl}  + \delta_{\rm ik} \delta_{\rm
  jl} + \delta_{\rm il} \delta_{\rm jk})  
\end{equation}
where $\sigma_{\rm T}^2 = S_2 \equiv \sigma_0^2$, $\sigma_{\rm H}^2 = S_6
\equiv \sigma_2^2$, $\Gamma_{\rm TH} = S_4 \equiv  \sigma_1^2$,
$\delta_{\rm ij}$ is the Kronecker delta, and
\begin{equation}
S_{\rm n} = {1 \over 2 \pi^2} \int_0^{\infty} k^{\rm n}~P(k)~W^2(k) {\rm d}k 
\end{equation}
 
\begin{equation}
\label{eq_sig}
\sigma_{\rm j}^2 = {1 \over 2 \pi^2} \int_{0}^{\infty} k^{\rm
  2(j+1)}~P(k)~W^2(k)~{\rm d}k \equiv S_{\rm 2(j+1)}.  
\end{equation}
In the main text we prefer to use $\sigma_{\rm T}$ and $\sigma_{\rm H}$, rather than the more familiar $\sigma_0$ and $\sigma_2$, for their intuitive meaning. 
In particular, the subscript $T$ always indicates that a quantity is related to the shear field, while the subscript $H$ denotes a quantity linked to the Hessian of the density field.

The shear and density Hessian $T_{\rm ij}$ and $H_{\rm ij}$ are real symmetric tensors, so they are specified
by 6 components. Whenever necessary, we label those components with
the symbols $\alpha$ or $\beta$ to indicate the various couples, where
$\alpha, \beta =(1,1), (2,2), (3,3), (1,2), (1,3), (2,3)$. 
It is also useful to introduce the vectors ${\bf T}$  and ${\bf H}$, derived from the components of their corresponding tensors, i.e.
${\bf T} = (T_{11},T_{22},T_{33},T_{12},T_{13},T_{23})$ and ${\bf H} = (H_{11},H_{22},H_{33},H_{12},H_{13},H_{23})$.
The constrained eigenvalues of the matrix having components $T_{\alpha}|H_{\alpha}$ will be 
indicated with  $\zeta_1, \zeta_2, \zeta_3$.
As shown in Rossi (2012), the covariance matrix of the joint probability distribution of $ {\bf T}$ and $ {\bf H}$
is simply
\begin{equation}
 {\bf \sf V} = \begin{pmatrix}
  \langle T_{\alpha} T_{\alpha} \rangle &   \langle T_{\alpha} H_{\beta} \rangle \\
    \langle H_{\beta} T_{\alpha} \rangle &   \langle H_{\beta} H_{\beta} \rangle 
 \end{pmatrix} =
 {1 \over 15} \begin{pmatrix}
 \sigma^2_{\rm T} {\bf \sf A} & \Gamma_{\rm TH} {\bf \sf A} \\
  \Gamma_{\rm TH} {\bf \sf A} & \sigma^2_{\rm H} {\bf \sf A}
\end{pmatrix}
\end{equation}
where
\begin{equation}
\label{matrix_A}
 {\bf \sf A} = \begin{pmatrix}
  {\bf \sf B} & \oslash \\
  \oslash & {\bf \sf I}  
 \end{pmatrix},~
  {\bf \sf B} = \begin{pmatrix}
  3 & 1 & 1\\
  1 & 3&1\\
  1&1&3 
 \end{pmatrix}
 \end{equation}
with ${\bf \sf I}$ a ($3\times3$) identity matrix and $\oslash$ a ($3 \times 3$) null matrix. 
Finally,  an important `spectral parameter' often used here
is the `reduced' correlation:
\begin{equation}
\gamma = \Gamma_{\rm TH}/\sigma_{\rm T} \sigma_{\rm H} = {\sigma_1^2 \over \sigma_0 \sigma_2},
\label{tilde_corr}
\end{equation}
which plays a crucial role in peaks theory (i.e. Bardeen et al. 1986). 
If Gaussian filters are used in (\ref{eq_sig}), then our $\gamma$
is the same as the one introduced in Bardeen et al. (1986) -- specified by their
Equation (4.6a). Note that in the main text we also define $\eta = \gamma \sigma_{\rm T} / \sigma_{\rm H}$;
if one adopts reduced variables (i.e. $T_{\rm \alpha}$ and $H_{\rm \alpha}$ normalized by their corresponding rms values $\sigma_{\rm T}$ and $\sigma_{\rm H}$), clearly $\eta \equiv \gamma$.


\section{Invariants from the conditional formulas}
\label{insights}


The algebra presented in Section \ref{pdesa}
allows one to gain  more insights into the 
joint conditional distribution of eigenvalues in the peak/dip picture (Equation \ref{doro_inter_extended}). 
For simplicity, in what follows we consider `reduced' variables, so that the various components of $\bf T$ and $\bf H$
are now normalized by their corresponding rms values ($\sigma_{\rm T}$ and $\sigma_{\rm H}$, respectively). With some abuse of notation, we
omit the tilde symbol (used instead in Rossi 2012) to distinguish between normalized and unnormalized quantities.
It is then possible to characterize and study the properties of the first few elementary symmetric functions of degree $n$ 
for the density Hessian, the shear tensor, and the conditional shear tensor -- 
along the lines of Weyl (1948), Doroshkevich (1970), Sheth \& Tormen (2002) and Desjacques (2008).
In particular, it is direct to note that, using Equation (\ref{H_gauss}) and considering six independent Gaussian random variates $y_{\rm i}$ ($i=1,6$) represented by the six-dimensional vector ${\bf y}$, 
one obtains that the first two classical invariants
\begin{eqnarray}
h_1 &=& H_{11}    +H_{22} +H_{33}  = - y_1\\
 h_3^2 &=& h_1^2 - 3 h_2 = {1 \over 5} (y_2^2 + y_3^2 +y_4^2+y_5^2+y_6^2)
\end{eqnarray}
are independent. This fact implies that
\begin{equation}
p({\bf H})  =  
{e^{-h_1^2/2} \over \sqrt{2 \pi}} 
{15^3 \over 8 \sqrt{10} \pi^{5/2}}  e^{-5 h_3^2/2}
\equiv p(h_1) p(h_3), 
\end{equation}
hence $p({\bf H})$ is the product of two independent distributions,  
where in particular  $p(h_1)$ is a one-dimensional Gaussian with mean zero and unity variance.
Note also that $p({\bf H}) {\rm d} {\bf H} = p(\bf y) {\rm d}{\bf y}$, where
$p ({\bf y}) \equiv  \prod_{\rm i=1}^6 g_{\rm i}$ is simply the product of six independent one-dimensional zero mean unit variance Gaussians $g_{\rm i}$.

Similarly, for the shear tensor ${\bf T}$, one obtains that
\begin{eqnarray}
k_1 &=& T_{11}    +T_{22} +T_{33}  = - z_1\\
 k_3^2 &=& k_1^2 - 3 k_2 = {1 \over 5} (z_2^2 + z_3^2 +z_4^2+z_5^2+z_6^2)
\end{eqnarray}
are also independent, with
$z_{\rm i}$ ($i=1,6$) other 
 six Gaussian random variates represented by the six-dimensional vector ${\bf z}$.
Therefore, 
\begin{equation}
p({\bf T}) =  
{e^{-k_1^2/2} \over \sqrt{2 \pi}}
{15^3 \over 8 \sqrt{10} \pi^{5/2}}  e^{-5 k_3^2/2} \equiv p(k_1) p(k_3)
\end{equation}
with $p(k_1)$ a one-dimensional Gaussian distribution. 
Note again that $p({\bf T}) {\rm d} {\bf T} = p({\bf z}) {\rm d}{\bf z}$, where
$p ({\bf z}) \equiv  \prod_{\rm i=1}^6 g_{\rm i}$ is the product of six independent one-dimensional zero mean unit variance Gaussians $g_{\rm i}$.

Following the previous logic, one would naturally expect that the quantities $K_1$ and
 $K_3^2 = K_1^2 - 3 K_2 $, defined in the main text (see Equation \ref{K_def}),
should also be independent.
Indeed, it is direct to obtain that (Ravi Sheth, private communication):
 \begin{eqnarray}
K_1 &=&  = -( m_1 - \gamma y_1 ) =  - \sqrt{1-\gamma^2}~l_1 \\
 K_3^2 &=& K_1^2 - 3 K_2 \nonumber \\ 
 &=& {1 \over 5}[ (m_2-\gamma  y_2)^2+(m_3-\gamma y_3)^2+(m_4-\gamma y_4)^2+(m_5-\gamma y_5)^2+(m_6-\gamma y_6)^2  ] \nonumber \\
   &= & {(1-\gamma^2) \over 5}~(l_2^2 +l_3^2 +l_4^2 +l_5^2 +l_6^2), 
\end{eqnarray}
where $l_{\rm i}$ ($i=1,6$) are  other six independent Gaussian distributed variates with mean zero and unity variance,
while $m_{\rm i}$ ($i=1,6$) are six Gaussian distributed variates with shifted mean $\gamma y_{\rm i}$ and reduced variance $(1-\gamma^2)$, i.e. $m_{\rm  i} = \gamma y_{\rm i}  + \sqrt{1-\gamma^2} ~ l_{\rm i} $.
Hence, the joint conditional distribution of eigenvalues in the peak/dip picture (Equation \ref{doro_inter_extended}) can be written as
the product of two independent
distributions as follows:
\begin{equation}
p ({\bf T}|{\bf H},\gamma) = {e^{-K_1^2/[2(1-\gamma^2)]  } \over \sqrt{2 \pi (1-\gamma^2)}} {15^3 \over 8 \sqrt {10} \pi^{5/2}} { e^{-5 K_3^2/[2(1-\gamma^2)] }    \over (1-\gamma^2)^{5/2}}
\equiv p(K_1|\gamma) p(K_3|\gamma).
\label{doro_inter_extended_more}
\end{equation}
This latter expression clearly shows that the distribution of constrained $K_1 \equiv \Delta_{\rm T|H}$ is independent
of the distribution of the constrained angular momentum $K_3^2$, where
in particular $p(K_1|\gamma)$ is a Gaussian with zero mean and variance given by $(1-\gamma^2)$, while $p(K_3)$ is a chi-square distribution with five 
degrees of freedom.
Once again, note that $p({\bf T}|{\bf H},\gamma) {\rm d} ({\bf T}|{\bf H}) = p({\bf m}|{\bf y},\gamma) {\rm d}({\bf m}|{\bf y})$, where now
\begin{equation}
p ({\bf m}|{\bf y},\gamma) \equiv  \prod_{\rm i=1}^6  { e^{(m_{\rm i} -\gamma y_{\rm i})^2 /2 (1-\gamma^2)} \over \sqrt{2 \pi (1-\gamma^2)}}  \equiv   \prod_{\rm i=1}^6 t_{\rm i};
\end{equation}
namely, $p({\bf m}|{\bf y},\gamma)$ is now the product of six independent one-dimensional Gaussians $t_{\rm i}$
with shifted mean $\gamma y_{\rm i}$ and reduced variance $(1-\gamma^2)$ -- represented by the six-dimensional vector ${\bf m}|{\bf y}$.


\label{lastpage}


\end{document}